\documentclass[12pt,preprint]{aastex}
\usepackage{color}

\def\kms{{\rm\,km\,s^{-1}}}
\def\kmskpc{{\rm\,km\, \,s^{-1} \, {kpc}^{-1}}}

\def\deg{{^\circ}}

\def\kpc{{\rm\,kpc}}

\def\mathnew{\mathsurround=0pt}   
\def\simov#1#2{\lower .5pt\vbox{\baselineskip0pt  
    \lineskip-.5pt\ialign{$\mathnew#1\hfil##\hfil$\crcr#2\crcr\sim\crcr}}}

\def\'#1{\ifx#1i{\accent"13\i}\else{\accent"13#1}\fi}    
  
\def\et{et~al. }     
 
\received{}
\begin{document}

\shorttitle{Pitch Angle Restrictions in Normal Spiral Galaxies}  
\shortauthors{P\'erez-Villegas et al. 2012}

\title{Stellar Orbital Studies in Normal Spiral Galaxies I: Restrictions to the Pitch Angle}

\author{A. P\'erez-Villegas, B. Pichardo, E. Moreno}     
 
\affil{Instituto de Astronom\'ia, Universidad Nacional
Aut\'onoma de M\'exico, A.P. 70-264, 04510, M\'exico, D.F.; Universitaria, D.F., M\'exico; \\barbara@astro.unam.mx}

\begin{abstract}
We built a family of non-axisymmetric potential models for normal
non-barred or weakly-barred spiral galaxies as defined in the simplest classification of galaxies:
the Hubble sequence. For this purpose a three-dimensional
self-gravitating model for spiral arms PERLAS is superimposed to
the galactic axisymmetric potentials. We analyze the stellar dynamics varying 
only the pitch angle of the spiral arms, from 4$\deg$ to 40$\deg$, for an Sa
galaxy, from 8$\deg$ to 45$\deg$, for an Sb galaxy, and from 10$\deg$
to 60$\deg$, for an Sc galaxy. Self-consistency is indirectly tested
through periodic orbital analysis, and through density response
studies for each morphological type. Based on ordered behavior,
periodic orbits studies show that for pitch angles up to approximately
$15\deg$, $18\deg$, and $20\deg$ for Sa, Sb and Sc galaxies,
respectively, the density response supports the spiral arms
potential, a requisite for the existence of a long-lasting
large-scale spiral structure. Beyond those limits, the density response tends to
``avoid'' the potential imposed by mantaining lower pitch angles in the
density response; in that case the spiral arms may be
explained as transient features rather than long-lasting large-scale structures. In a second limit, from a phase space orbital study based on chaotic behavior, we found that for pitch angles larger than $\sim30\deg$, $\sim40\deg$ and $\sim50\deg$ for Sa, Sb, and Sc galaxies, respectively, chaotic orbits dominate all phase space prograde region that surrounds the periodic orbits sculpting the spiral arms and even destroying them. This
result seems to be in good agreement with observations of pitch angles
in typical isolated normal spiral galaxies.

\end{abstract}


\section{Introduction}                                                     
\label{sec:intro}    
When Hubble (1926, 1936) introduced his classification scheme of galaxies, he emphasized that the turning-fork diagram he obtained,
depicted the systematic variation of the morphological
characteristics, and the terms ``early type" and ``late type" referred
only to the relative position of a galaxy in this empirical sequence.
The sequence did not imply temporal evolution connections. Some
modifications have been introduced to this scheme over the years (de
Vaucouleurs 1959; Sandage 1961). In the case of isolated galaxies,
astronomers keep searching in this well-ordered sequence of galaxy
types for a clue to possible formation and evolutionary processes to
explain galactic morphology. It is not surprising then, that
morphology is frequently an underlying theme in the study of
galaxies. A few percent of all galaxies are unclassifiable, many of
these due to their unusual morphology produced by their interacting
nature. In this paper we focus on intrinsic dynamical processes of galaxies,
this refers to relatively isolated galaxies. 

Spiral galaxies are classified in the Hubble sequence based mainly on
two criteria: the pitch angle and the bulge to disk luminosity ratio. On Sandage's classification (Sandage 1975), three parameters to classify spiral galaxies are employed: the bulge to disk ratio, the pitch angle of the spiral arms, and the arms fragmentation into stars. These two classifications are
very similar. Then, early type spirals show small pitch angles with smooth structure and conspicuous central bulges. For late spirals, the arms are open and
flocculent, and the central bulges are smaller. Finally, Sb galaxies are
between the above two types.

Since the Hubble classification was introduced, astronomers are puzzled and searching correlations or
dependences between galactic parameters and morphological
types. Holmberg (1958) and S\'ersic (1987), compiled and analyzed
photometric data (integrated magnitudes, colors and diameters) for
hundreds of galaxies, and found a correlation between the Hubble type and galaxy colors. The mass ratio
$M_{H_2} / M_{Tot}$ decreases with type from Sa to Sd (Sage 1993;
Roberts \& Haynes 1994). The total mass decreased as the Hubble type
varied from Sa to Sc (Pi\c smi\c s \& Maupom\'e 1978; Maupom\'e, Pi\c
smi\c s \& Aguilar 1981; Roberts \& Haynes 1994). Another interesting
correlation is found in the maximum value of rotation velocity: values for Sa
galaxies are higher than Sc galaxies (Rubin et al. 1985; Sandage 2000;
Sofue \& Rubin 2001); although, the maximum rotation velocity presents
a large scatter. Along the Hubble sequence, spiral galaxies disks tend
to be thinner (Ma 2002). Some correlations are also found with the
pitch angle, like the one related with the central supermassive black hole in
spiral galaxies whose mass seems to decrease with the pitch angle (Seigar et
al. 2008; Shields et al. 2010); spiral galaxies with higher rotational velocity also have tighter spirals (Kennicutt 1981; Savchenko \& Reshetnikov 2011);
in the same manner, it seems that for disks with lower surface
densities and lower total mass-luminosity ratios, pitch angles tend to
be larger (Ma 2002).

Regarding specifically to spiral arms morphology, studies of this type
in galaxies, started earlier than the Hubble
classification. Von der Pahlen (1911), Groot (1925), Danver (1942), and
Kennicutt (1981) found that spiral arms in galaxies are well fitted with logarithmic
helices. Kennicutt (1981) concluded that the pitch angle in early type galaxies tends to be smaller than in late type galaxies, but there is a large scatter in the pitch angle within each morphological type, concluding that the ideal Hubble classification is not closely followed by real galaxies (Kennicutt 1981; Ma \et 2000; Davis \et 2012).

In the Lin \& Shu (1964) spiral density wave theory, galactic spiral
arms are modeled as a periodic perturbation to
the axisymmetric background disk's potential. This is known as the
tight-winding approximation (TWA) for small pitch angles, or WKB
approximation (Wentzel-Kramers-Brillouin approximation of quantum mechanics -Binney \& Tremaine 1994). The solution for bisymmetric spiral arms provided by the TWA takes the form,

\begin{eqnarray} \label{cosine}
\Phi(R,\phi)= f(R)cos[2\phi + g(R)] .
\end {eqnarray}
where the function $f(R)$ is the amplitude of the perturbation, and $g(R)$
represents the geometry of the spiral pattern. The amplitude function
$f(R)$ given by Contopoulos \& Grosb\o l (1986) and $g(R)$ of the form
presented by Roberts, Huntley \& van Albada (1979) are commonly used. Most
orbital studies in spiral galaxies theory, have employed a spiral
potential of the form in equation \ref{cosine}. However, the majority
of disk galaxies possesses strong spiral structures, making clear that
this large scale structure is an important non-axisymmetric component
of galaxies that deserves an effort to model it beyond a simple perturbing term.
In this work we have chosen a more physically and observationally
motivated model of the spiral arms for disk galaxies, the one called
PERLAS from Pichardo \et (2003), based on a three-dimensional model
mass distribution, which allows a more
detailed representation of spiral arms.

Concerning to large scale structures in galaxies, these seem to be related mainly to ordered orbital behavior (Patsis 2008). This is the case for example, of grand design galaxies, where non-axisymmetric structures are composed by material librating around the family of periodic orbits known as
X$_1$ (Contopoulos 2002). In spiral arms, where the X$_1$ periodic orbits dominate, the flow of material through the arms can be described as a
``precessing ellipses" flow. This is supported by many observational and theoretical
studies (Patsis, Contopoulos \& Grosb\o l 1991). Spiral galaxies, with their
thin disks and smooth spiral arms may seem to be dominated by ordered
motion, but deeper orbital studies show that chaos may be significant in spiral galaxies
(Contopoulos 1983,1995; Contopoulos \et 1987; Grosb\o l 2003; Voglis,
Stauropoulos \& Kalaptharakos 2006; Contopoulos \& Patsis 2008; Patsis
\et 2009), and that chaotic orbits surrounding stable periodic orbits
could also support the spiral arms. However, spiral arms and bars, for example, are
not expected to originate out of only chaotic orbits. Patsis \&
Kalapotharakos (2011) call {\it ordered} spirals those that have as
building block a set of stable periodic orbits; on the other hand,
they call {\it chaotic } spirals those that they believe are
constituted mainly from stars in chaotic motion.

Currently, discussion about the nature of spiral arms as a
long-lasting or transient structure is ongoing in the
field. Theoretical studies have demonstrated that spiral arms might be
rather transient structures (Goldreich \& Lynden-Bell 1965; Julian \&
Toomre 1966; Sellwood \& Carlberg 1984; Foyle \et 2011; Sellwood 2011;
P\'erez-Villegas \et 2012; Fujii \& Baba 2012; Kawata \et
2012). However, if spiral arms are smooth, weak and/or with small
pitch angles, the spiral structure could be long-lasting
(P\'erez-Villegas \et 2012, that study is referred to late type
spirals).

In a previous paper (P\'erez-Villegas \et 2012), we found two clear limits for
the pitch angle in late (Sc) spiral galaxies, one that sets a maximum limit for which
steady spiral arms are plausible (based on periodic orbital studies), beyond which, a transient
nature for the arms is proposed; the second limit is for which spiral arms are so opened that orbital chaos dominates completely. In this second paper, we extend our studies to early (Sa) and intermedium (Sb) galaxies to see if these limits found in late spirals are also followed by the
rest of spirals in the Hubble sequence and if the values of those limits match with observations. We isolate the effect of the pitch angle, that represents the least restricted parameter (i.e. the one with more spread values going from 4$\deg$ to 50$\deg$), and for that reason the one with likely more effect on the stellar dynamics. We employ for this purpose, realistic values for the rest of the parameters that identify approximately typical Sa, Sb and Sc galaxies. The rest of the structural parameters such as angular speed, perturbation strength (spiral arms mass), etc., are better restricted in the sense that the ranges for those values are tighter, measured both by observations and/or by self-consistent models. Specific calculations regarding these parameters (angular speed, spiral arms strength, and axisymmetric components) will be discussed in a forthcoming paper.

This paper is organized as follows. The galactic potential and methodology are described in Section \ref{model}. Our results: two pitch angle restrictions in galaxies, the
first based on ordered orbital behavior, and the second based on
chaotic behavior,  are presented in Section \ref{results}. Finally,  we present a discussion and
our conclusions in Section \ref{conclusions}.

\section{Methodology and Numerical Implementation}\label{model}

We have constructed a family of models for the potential of normal
spiral galaxies (Sa, Sb and Sc), as classified by Hubble (1926), and
also based on recent observational parameters taken from the literature. For
this purpose we solved numerically the equations of motion for
stars in an axisymmetric potential plus a spiral arm potential. The
main methods to study stellar dynamics that we use in this work are
periodic orbital analysis and Poincar\'e diagrams.

\subsection{Models for Normal Spiral Galaxies}

The most common method to model spiral arms is a two-dimensional
bisymmetric local potential approximated by a cosine function (based on
the solution for the TWA). It assumes that spiral arms are smooth,
self-consistent perturbations to the axisymmetric potential. In this
regime, self-consistency can not be assured if a little larger pitch
angles and/or slightly larger amplitudes to the spiral arms than the
ones assumed for the TWA, are imposed. 

To model typical spiral arms in galaxies (i.e. pitch angles larger than about 6$\deg$, or mass of the spiral arms larger than 1$\%$ of the disk mass), one goes readily far away
from the TWA self-consistency limits. Spiral arms in real galaxies,
are complicated three-dimensional gigantic structures, far intricate to be approximated with a function as simple
as a cosine. To test self-consistency in these type of potentials, we analyzed ordered and chaotic orbital behavior and specifically the construction of periodic orbits. Between a cosine potential and a potential based on mass distribution, the differences are not
negligible at all, specially, when we are dealing with chaos.  

To this purpose, we employ the spiral arms potential called PERLAS from Pichardo \et
(2003), this potential is formed by individual  potentials of oblate
inhomogeneous spheroids lying along the logarithmic spiral
locus given by Roberts, Huntley, \& van Albada (1979),

 \begin{eqnarray} \label{locus}
g(R)=-\Big( \frac{2}{N\tan_{i_p}} \Big)\ \ln [1+(R/R_s)^N] ,
\end {eqnarray}
with $i_p$ the pitch angle. $R_s$ marks the galactocentric position
where the spiral arms begin, and $N$ is a constant that gives the
shape to the starting region of the spiral arms (we set it to 100; for
details see Pichardo \et 2003).

This bisymmetric self-gravitating potential is more
realistic since it is based on a three-dimensional density distribution, providing a much more complicated function for the
gravitational potential, unlike a two-dimensional local arm like the
cosine potential. PERLAS is observationally motivated; comparison with other theoretical models have been already published (Pichardo \et 2003; Martos
\et 2004; Antoja \et 2009, 2011). We have tested approximately the self-consistency of the model, through the reinforcement by the stellar orbits (Patsis \et 1991; Pichardo \et 2003).

\begin{figure}

\includegraphics[width=1\textwidth]{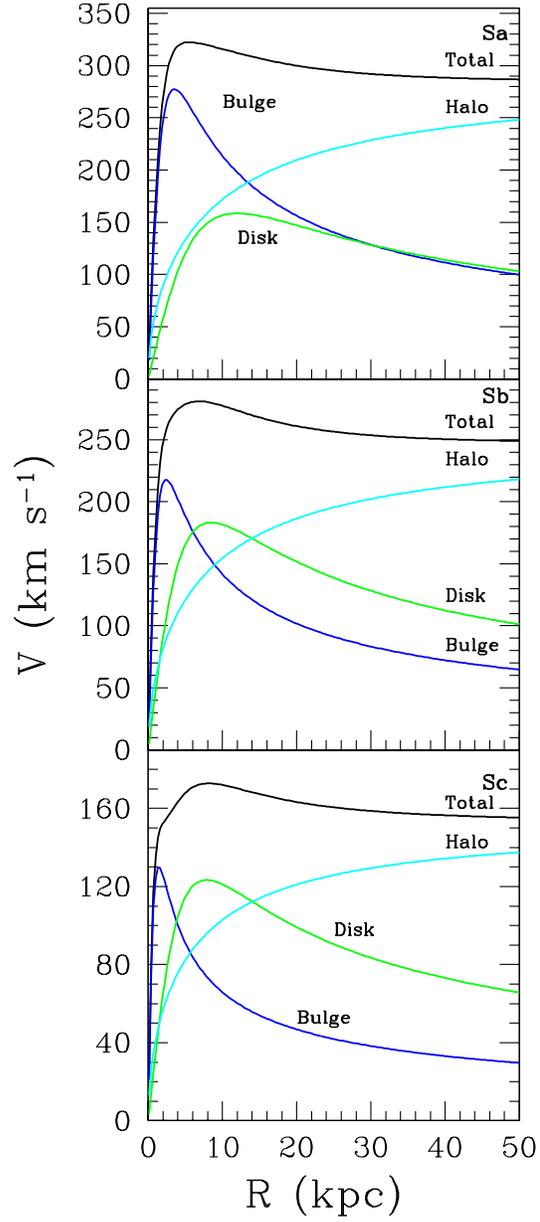}
\caption{Rotation curves for Sa (top panel), Sb (middle panel) and Sc
  (bottom panel) galaxies.}
\label{rotation_curve}
\end{figure}

Parameters used to fit normal spiral
galaxies (Sa, Sb and Sc), with references are presented in Table \ref{tab:parameters}.
The spiral arms potential is superimposed on an axisymmetric background,
described by a massive halo (Allen \& Santill\'an 1991), and a Miyamoto-Nagai (1975) disk and bulge. We fitted the rotation curve based on observational data from literature of 
the masses of the disk and bulge $vs.$ disk mass (depending on morphological type).  With this information, we derived a halo mass using a typical maximum velocity in the rotation curve for each type. In Figure \ref {rotation_curve}, we show the resulting rotation curves.

\begin{figure}
\includegraphics[width=1\textwidth]{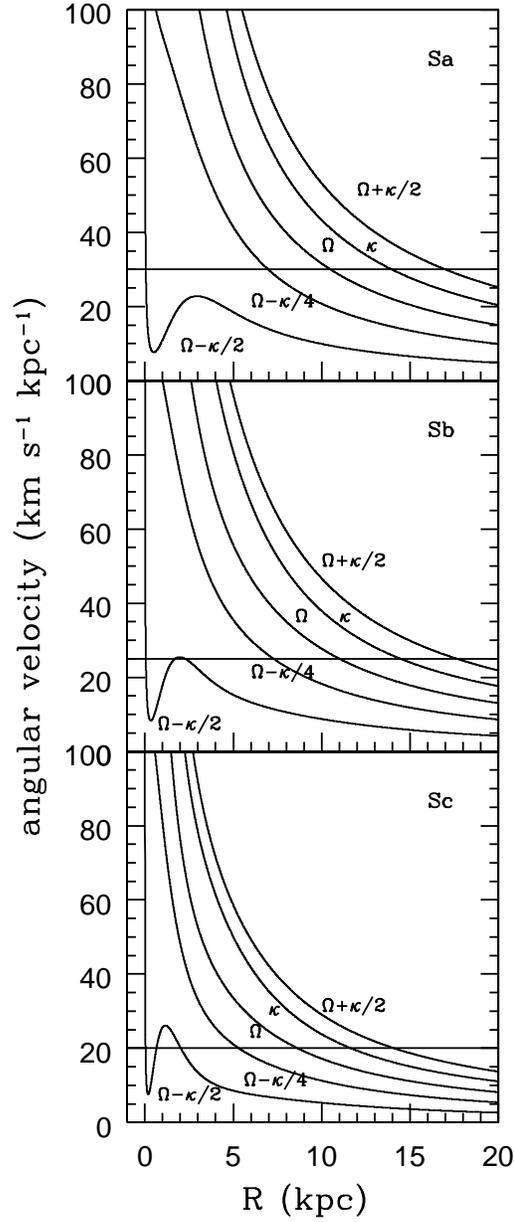}
\caption{Resonance diagrams for Sa (top panel), Sb (middle panel) and
  Sc (bottom panel) galaxies. Horizontal lines denote the spiral
  pattern angular speed: $|\Omega_p|= 30, 25 \, {\rm and} \, 20
  \kmskpc$ for an Sa, an Sb and an Sc, respectively.}
\label{resonance}
\end{figure}

\begin{figure}
\includegraphics[width=1\textwidth]{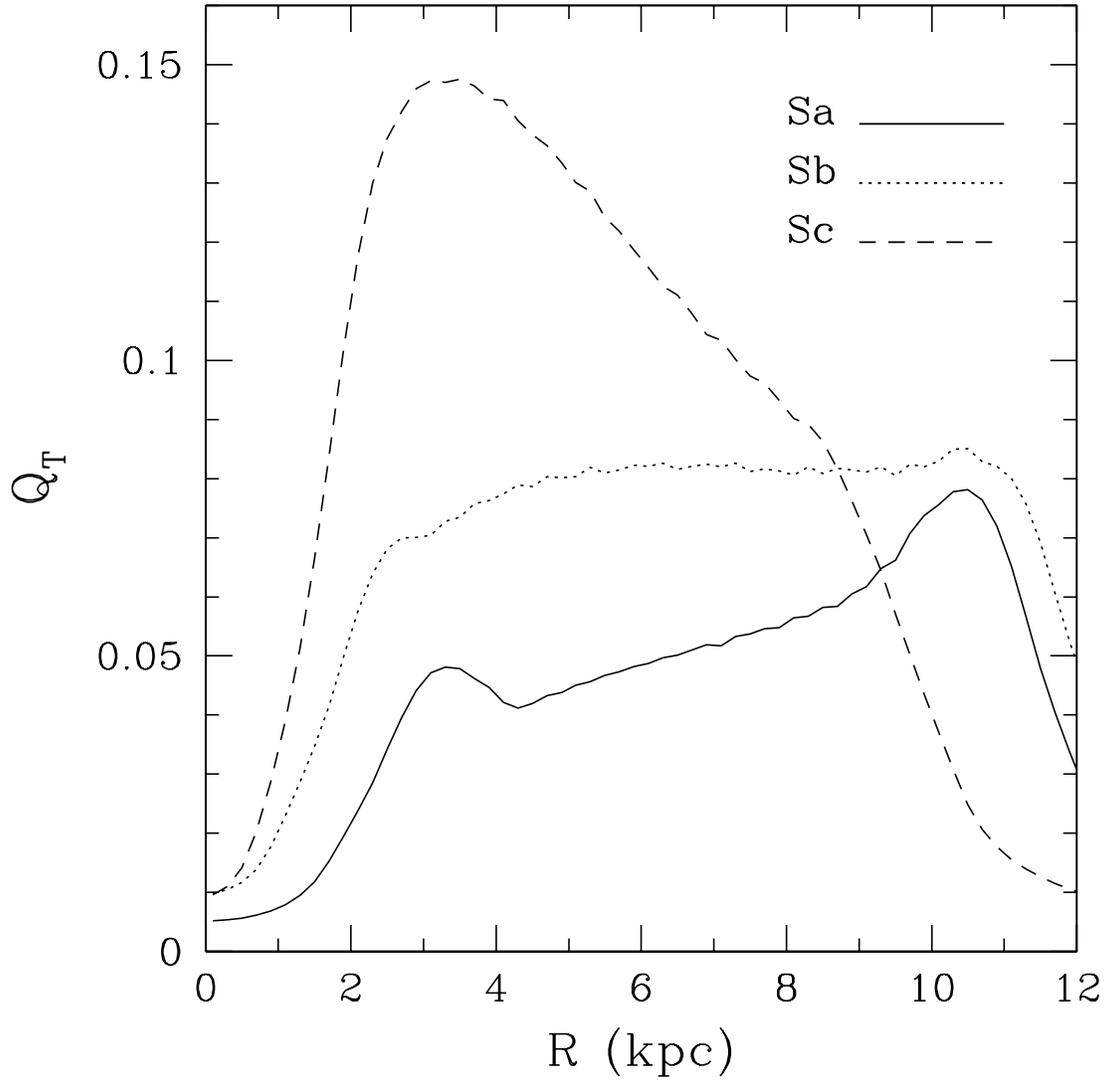}
\caption{ $Q_T$ parameter (equation \ref{q_max}) for an Sa (solid line), Sb (dotted line), and Sc (dashed line) galaxy, where the pitch angles are $30\deg$, $40\deg$ and $50\deg$ for an Sa, Sb and Sc galaxy, respectively.}
\label{QT}
\end{figure}

\begin{figure}
\includegraphics[width=1\textwidth]{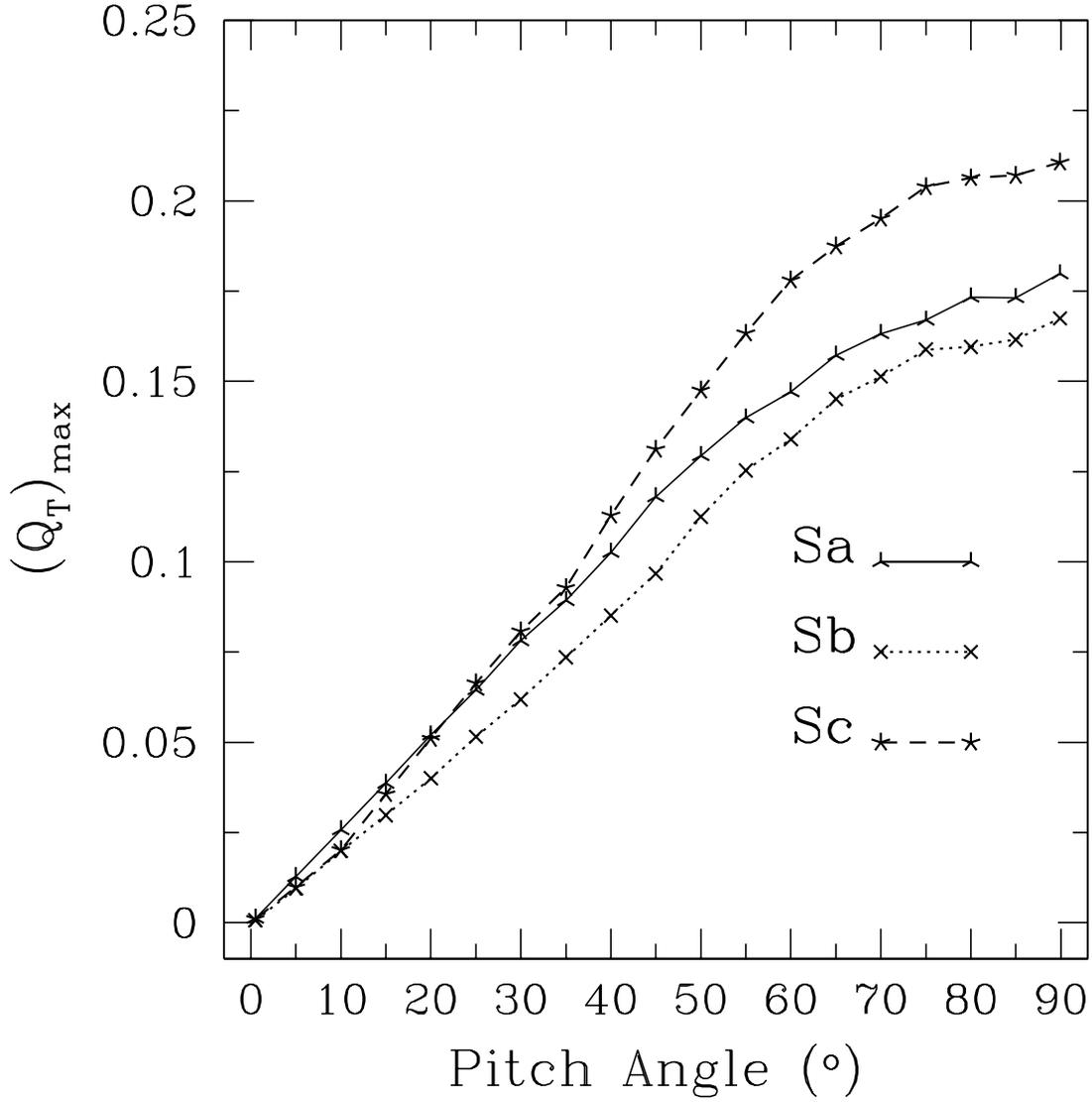}
\caption{Maximum value, (Q$_{\rm T})_{\rm max}$, of the
  parameter Q$_T(R)$ (maximum relative torques) vs. pitch angle of the
  spiral arms. Solid line gives (Q$_{\rm T})_{\rm max}$
  for an Sa galaxy, dotted line 
  for an Sb galaxy and dashed line for an Sc galaxy. }
\label{parameterQ}
\end{figure}

Theoretical studies of orbital self-consistency in normal spiral
galaxies show that weak spiral arms end at corotation and strong
spiral arms end at the 4/1 resonance (Contopoulos \& Grosb\o l 1986,
1988; Patsis, Contopoulos \& Grosb\o l 1991). Then, the spiral arms
limits depend on the position of the Inner Lindblad Resonance (ILR),
for the inner part, and corotation for the outer part of the spirals
(in our models we place the spiral arms between these two
resonances). Consequently, the angular speed of the spiral arms,
$\Omega_p$, sets the corresponding value of the beginning and the end of
these arms. Measuring spiral arms angular velocities
observationally is not an easy task. We know however that on average,
earlier type galaxies tend to be more massive, therefore Sa 
galaxies rotate faster than Sc galaxies. On the other hand, considering the few observational studies of angular speeds from literature, these speeds seem to present a small range of difference from Sa to Sc, ranging between 35 and 15 $\kmskpc$. We employed for the axisymmetric models mass average values for the different components. With a clockwise rotation for the disk, we assume an angular velocity of
$\Omega_p = -30, \, -25$, and $-20 \, \kmskpc$, for the spiral arms in
Sa, Sb, and Sc galaxies, respectively (see Table \ref{tab:parameters}
for references). In Figure \ref{resonance} we show resonance diagrams
for our galactic models. We take a fairly good approximation to the spiral arms mass, of 3\% of the total disk mass (Pichardo \et 2003), independently of the Hubble type. Additionally, to assure the spiral mass we fixed to our
models is within observational limits, we have employed the parameter $Q_T$ (Combes \& Sanders 1981). This parameter has been
implemented in studies of bars and spiral arms (Buta \& Block 2001; Laurikainen \&
Salo 2002; Buta \et 2004; Laurikainen \et 2004; Block \et 2004; Vorobyov 2006; Kalapotharakos \et 2010) to measure the
strength of large scale non-axisymmetric structures in galaxies. The parameter $Q_T$ is
defined as

\begin{eqnarray} \label{q_max}
Q_{\rm T}(R)=F_{T}^{\rm max}(R)/|\langle F_{\rm R} (R) \rangle|, 
\end {eqnarray}
where F$^{\rm max}_{ T}({\rm R})$ =$|\left(\frac{1}{R} \ \partial\Phi
({\rm R},\theta)/\partial \theta\right)|_{\rm max}$, represents the
maximum amplitude of the tangential force at radius R, and
$\langle$F$_{\rm R}$(R)$\rangle$, is the mean axisymmetric radial
force at the same radius, derived from the ${\rm m}=0$ Fourier component of
the gravitational potential. In Figure \ref{QT} we show an example of the behavior of the parameter $Q_{\rm T}(R)$ given by equation \ref{q_max}, for an Sa (solid line), Sb (dotted line) and Sc (dashed line) galaxies. In these three cases we have set the pitch angles to their maximum values permitted before chaos destroys all the periodic orbits support ($30\deg$, $40\deg$, $50\deg$, for Sa, Sb, and Sc galaxies, respectively). Note how even for cases with high pitch angles, the parameter $Q_{\rm T}(R)$ keeps always lower than the observed maximum values in galaxies (Buta \et 2005). In Figure \ref{parameterQ} we present the
maximum value of the parameter $Q_T$ for each type of galaxy as we increase the pitch angle  from 0$^\deg$ to 90$^\deg$. Values up to
0.4 for the $Q_T$ parameter are consistent with observed spirals (Buta \et 2005).

\begin{deluxetable}{lcccl}
\tablecolumns{5}
\tabletypesize{\small}
\tablewidth{0pt}
\tablecaption{Parameters of the Galactic Models}
\tablehead{{Parameter} &\multicolumn{3}{c}{Value}& {Reference}}
\startdata

 &\multicolumn {3}{c}{\it Spiral Arms}  \\
\hline
  &Sa&Sb&Sc &\\
\hline
locus             & \multicolumn{3}{c}{Logarithmic } & 1,9\\
arms number       & \multicolumn{3}{c}{2} & 2\\
pitch angle ($\deg$)       & 4-40& 8-45& 10-60& 3,7 \\
M$_{arms}$/M$_{disk}$& \multicolumn{3}{c}{3\%} &  \\
scale-length (disk based) ($\kpc$)    & 7&5& 3  & 4,5\\
pattern speed ($\kmskpc$) &-30 &-25 & -20& 1,6 \\
 (clockwise) &&&&\\
ILR position ($\kpc$)    &3.0 &2.29 & 2.03& \\
corotation position (CR) ($\kpc$)  & 10.6& 11.14 & 8.63 & \\
inner limit   ($\kpc$)    & 3.0& 2.29& 2.03& $\sim$ILR position\\
&&&&based\\
outer limit ($\kpc$)    & 10.6& 11.14& 8.63 & $\sim$CR position\\ 
&&&&based\\
\hline
 &\multicolumn {3}{c}{\it Axisymmetric Components}  \\
\hline
M$_{disk}$ / M$_{halo}$ $^{\rm 1}$&0.07 &0.09 & 0.1  &  4,8  \\
M$_{bulge}$ / M$_{disk}$ & 0.9& 0.4& 0.2 & 5,8 \\
Rot. Curve (V$_{max}$) ($\kms$)& 320&250 &  170& 7  \\
M$_{disk}$ (M$_\odot$)   &$12.8\times10^{10}$ & $12.14\times10^{10}$& $5.10 \times10^{10}$& 4  \\
M$_{bulge}$ (M$_\odot$)  & $11.6\times10^{10}$& $4.45\times10^{10}$& $1.02 \times10^{10}$& $M_{disk}/M_{bulge}$ \\
&&&&based  \\
M$_{halo}$ (M$_\odot$)     &$1.64\times10^{12}$ &$1.25 \times10^{12}$& $4.85 \times10^{11}$& $M_{disk}/M_{halo}$ \\
&&&& based  \\
Disk scale-length ($\kpc$) & 7&5 & 3 & 4,5\\
\hline
&\multicolumn {3}{c}{\it Constants of the Axisymmetric Components $^{\rm 2}$}  \\
\hline
Bulge (M$_{bulge}$, b$_1$)$^3$& 5000, 2.5&2094.82, 1.7&440, 1.0&\\
Disk (M$_{disk}$, a$_2$, b$_2$)$^3$&5556.03, 7.0, 1.5&5232.75, 5.0, 1.0& 2200, 5.3178, 0.25&\\
Halo (M$_{halo}$, a$_3$)$^3$&15000, 18.0&10000, 16.0&2800, 12.0&
\label{tab:parameters}
\enddata

\tablenotetext{1} { Up to 100 kpc halo radius.} 
\tablenotetext{2} { In galactic units, where a galactic mass unit $= 2.32 \times10^7$ M$_\odot$ and a galactic distance unit = kpc.} 
\tablenotetext{3}{b$_1$, a$_2$, b$_2$, and a$_3$ are scale lengths.}

\tablerefs{ 1)~Grosbol \& Patsis 1998.
            2)~Drimmel \et 2000; Grosb\o l \et 2002.
            3)~Kennicutt 1981. 
            4)~Pizagno \et 2005
            5)~Weinzirl \et 2009.
            6)~Patsis \et 1991; Grosb\o l \& Dottori 2009; 
                 Egusa \et 2009; Fathi \et 2009.
	    7)~Brosche 1971; Ma \et 2000; Sofue \& Rubin 2001.
            8)~Block \et 2002.
            9)~Pichardo \et 2003.
}

\end{deluxetable} 

\subsection {Orbital Analysis}
For the orbital analysis we employed periodic orbits and Poincar\'e
diagrams. With this extensive phase space and configuration space orbital study, we are able to set two restrictions to one of the structural parameters of spiral arms: the pitch
angle. The motion equations are solved in the non-inertial reference
system of the spiral arms, and in Cartesian coordinates
$(x', y', z')$.

\subsubsection {Periodic Orbits}

Periodic orbits represent the simplest orbits of potentials in
general. They are also the most important orbits, because these are
followed by sets of non-periodic orbits, and even for chaotic regions confined between two quasiperiodic orbits,
librating around the periodic ones (i.e. forming tubes surrounding the
periodic orbits). In self-consistent systems, periodic orbits support
large-scale structures, such as bars and spiral arms, they are known as the ``dynamic backbone'' of potentials.

We computed between 40 and 60 periodic orbits for each Poincar\'e diagram, using the Newton-Raphson method. As a
first guess for the initial conditions in the calculation of the
periodic orbit the code provides the periodic orbit at a given radius
in the background axisymmetric potential. The orbits are launched from
$y'=0$, on the $x'$-axis $>0$ with $v'_x=0$ and $v'_y=v_c$, where
$v_c$ is the corresponding circular velocity given by,
 
\begin{eqnarray} \label{V_c}
v_c=\left(x' \left | \frac{d\Phi_0}{dx'} \right | \right)^{1/2},
\end {eqnarray}
and $\Phi_0$ is the axisymmetric potential. 

Periodic orbits close themselves in one or more periods; we use this
property to find them. In this manner, the Newton-Raphson method
searches a root based on the minimum distance and velocity between the final and initial points of the orbit after it completes a period. When the method finds a root, we obtain a periodic orbit. 

\subsubsection {Density Response} 

To calculate the density response to a spiral potential, we employ the method of Contopoulos \& Grosb\o l (1986), that quantifies the support with periodic orbits to the arms. The
method assumes that stars in circular orbits in an axisymmetric potential, rotating in the same direction of the spiral perturbation, will be trapped around the
corresponding periodic orbit in the presence of the spiral arms. To this purpose, we computed a large set of periodic orbits and calculated the density response along their
extension using the conservation of mass flux between any two
successive orbits. With this information we search the position of the density response
maxima along each periodic orbit. The locus of the obtained positions is compared with the imposed locus of PERLAS.

We calculated the average
density response around each one of these response maxima, taking a circular
vicinity with a radius of 500 pc. We then compared the density response
with the imposed density. The imposed density is the sum of the
axisymmetric disk density on the galactic plane and the central
density of the spiral arms.

\subsubsection {Poincar\'e Diagrams}

In Poincar\'e diagrams ordered orbits appear as invariant
one-dimensional curves; periodic orbits, on the other hand, draw a
finite set of dots. In phase space diagrams, chaotic
orbits appear as scattered sets of dots.

In the non-inertial frame, the effective potential on the galactic plane is
given by:

\begin{eqnarray} \label{pot_eff}
\Phi_{\rm {eff}} (x',y') = \Phi_{0} + \Phi_{sp}(x',y') -  \frac{1}{2} \Omega^{2}_{p}(x'^2+ y'^2) ,
\end {eqnarray}
 where $\Phi_0$ is the axisymmetric potential, $\Phi_{sp}$ is the
 potential of the spiral arms, and $\Omega_p$ is its angular
 velocity. The only known analytical integral of stellar motion in the
 non-inertial system of reference is the Jacobi constant, given by
 
\begin{eqnarray} \label{Ej}
E_J=\frac{1}{2}v'^{2}+ \Phi_{\rm {eff}},
\end {eqnarray} 
where $v'$ is the star velocity.  Poincar\'e diagrams are constructed
following the usual procedure. They present two regions, each one containing 50 orbits with
300 points each (corresponding to the number of periods), with a given prograde or retrograde sense of rotation, defined in the
galactic non-inertial frame. In our models, the spiral patterns move
in the clockwise sense. Therefore, the right side of the diagram (launching orbits with $x'>0, v'_y>0$) is the retrograde region, while the left side (with $x'<0, v'_y>0$),
is the prograde region.


\section{Results} \label{results}
We carried out an extensive orbital study with periodic orbital
analysis, density response and studies in the phase space to determine
whether limit values to different structural parameters of normal
spiral galaxies can be established.
 
We produced axisymmetric
potential models to simulate typical Sa, Sb and Sc spiral galaxies,
and superposed a spiral arms potential (PERLAS) to study the stellar orbital behavior in the disk, as
we changed the pitch angle for each galaxy type. Normal spiral galaxies
present a wide scatter in the pitch angle, ranging from $\sim$ 4$\deg$
to 50$\deg$. We have found two restrictions for the pitch angle in
normal spiral galaxies, the first based on ordered behavior and the
second based on chaotic behavior.

\subsection{Pitch Angle Restriction Based on Ordered Motion: long lasting or transient spiral arms} \label{order}
\begin{figure}[H]
\includegraphics[width=1\textwidth]{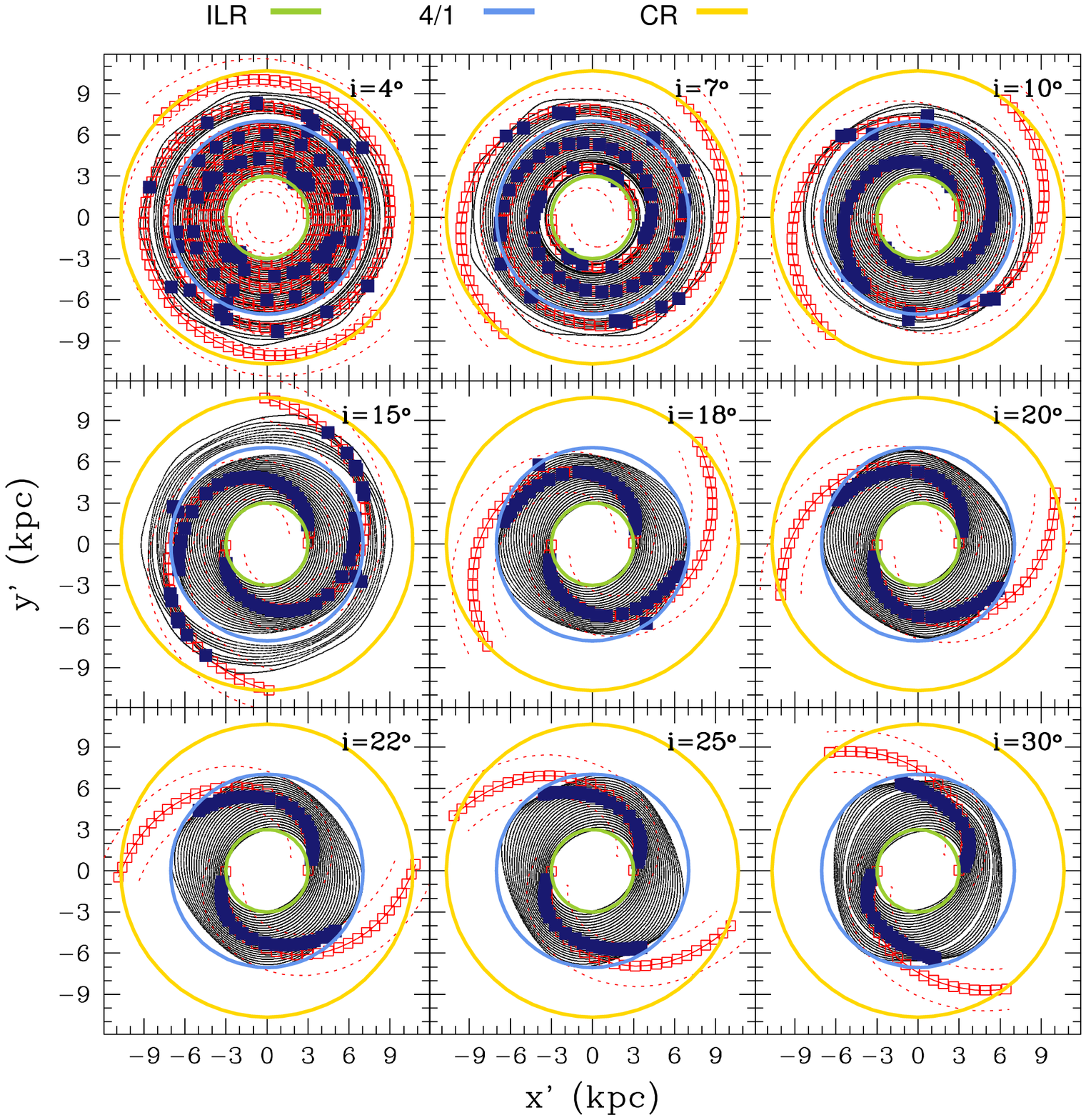}
\caption{We present 9 panels showing the periodic orbits, response
  maxima (filled squares), and the spiral locus (open squares), for
  the three-dimensional spiral model of an Sa galaxy (Table
  \ref{tab:parameters}), with pitch angles ranging from $4\deg$ to
  $30\deg$.}
\label{periodic_Sa}
\end{figure}

\begin{figure}[H]
\includegraphics[width=1\textwidth]{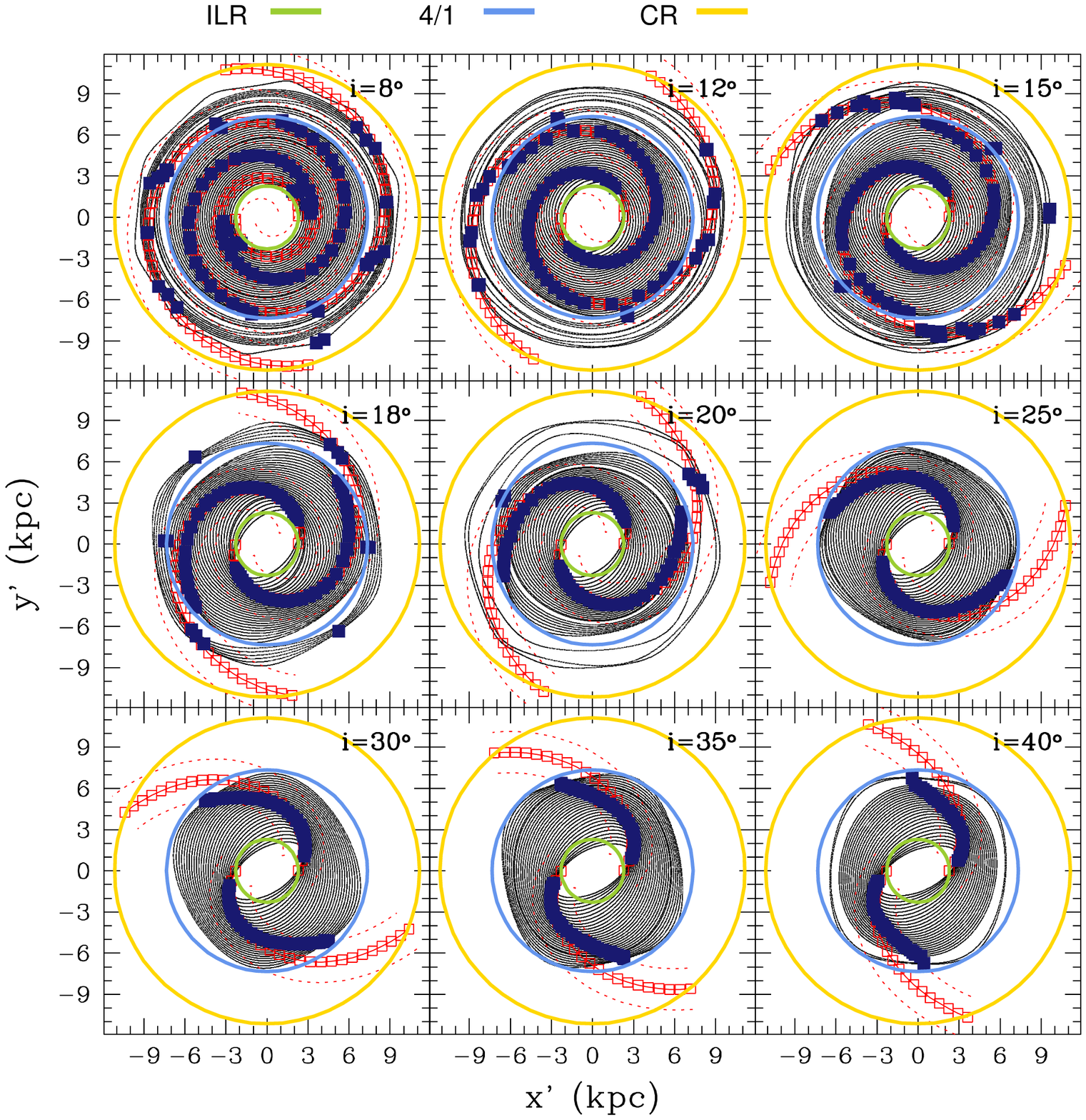}
\caption{We present 9 panels showing the periodic orbits, response
  maxima (filled squares), and the spiral locus (open squares), for
  the three-dimensional spiral model of an Sb galaxy (Table
  \ref{tab:parameters}), with pitch angles ranging from $8\deg$ to
  $40\deg$.}
\label{periodic_Sb}
\end{figure}

\begin{figure}[H]
\includegraphics[width=1\textwidth]{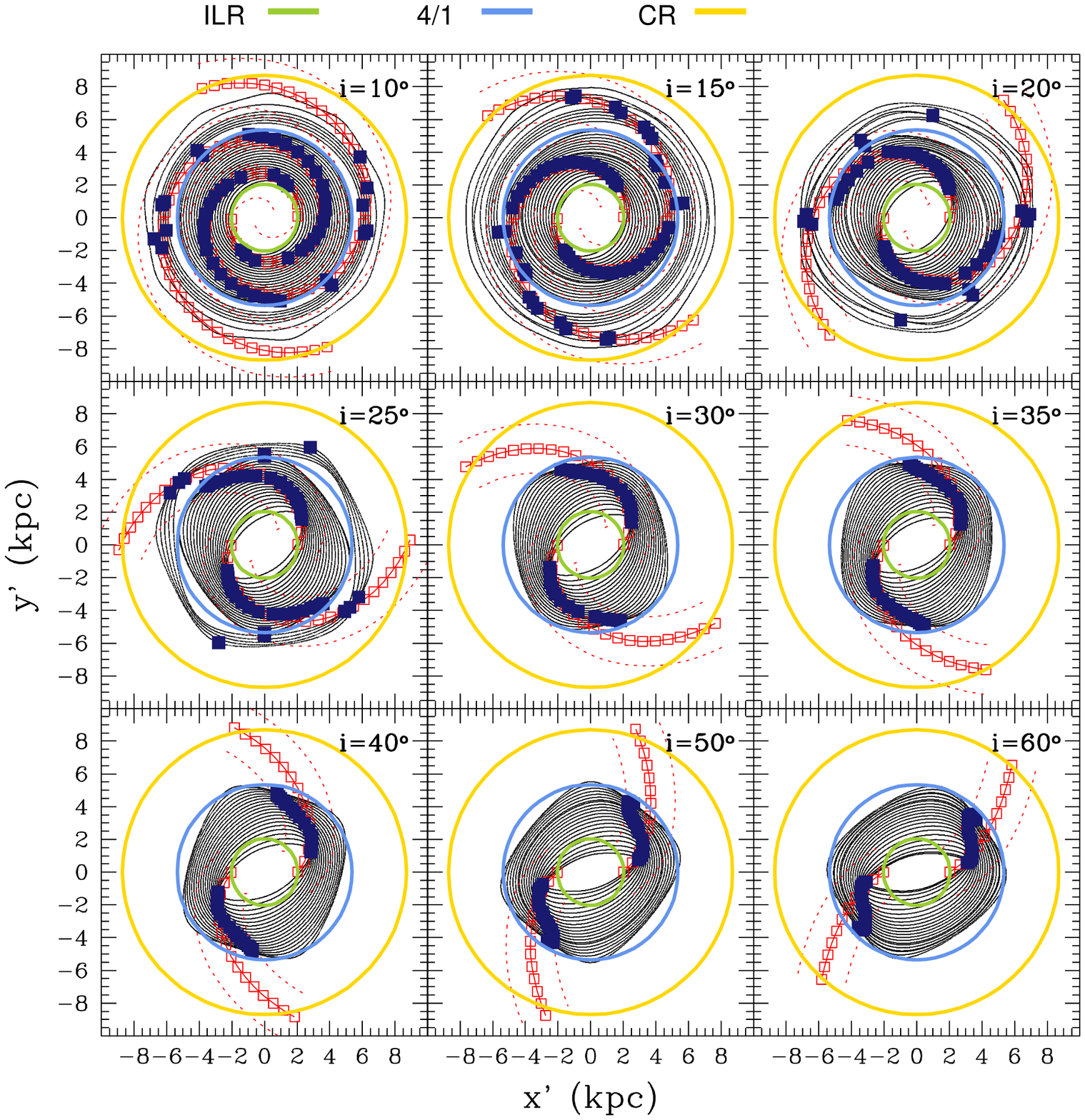}
\caption{We present 9 panels showing the periodic orbits, response
  maxima (filled squares), and the spiral locus (open squares), for
  the three-dimensional spiral model of an Sc galaxy (Table
  \ref{tab:parameters}), with pitch angles ranging from $10\deg$ to
  $60\deg$.}
\label{periodic_Sc}
\end{figure}

We estimated the self-consistency of each one of our models through the construction of periodic orbits. The 
existence of periodic orbits makes it more likely in steady conservative potentials, the support to long-lasting large-scale structures. We present a
periodic orbital study for each morphological type (see Figures
\ref{periodic_Sa} -- \ref{periodic_Sc}). In addition, we search the
position of the density response maxima along each periodic orbit. We
compared these positions with the center of the imposed spiral arms.
Figures \ref{periodic_Sa} -- \ref{periodic_Sc} show the response
maxima as filled squares, where the orbits crowd
producing a density enhancement, and the imposed spiral pattern are
represented by open squares. In these figures we present periodic
orbits for Sa, Sb and Sc galaxies (Figure \ref {periodic_Sa}, \ref {periodic_Sb}, 
and \ref{periodic_Sc}, respectively). We have
used the same axisymmetric background potential for each morphological
type, based on the parameters presented in Table \ref{tab:parameters}.
The pitch angle in these figures ranging from 4$\deg$ to 40$\deg$ for
an Sa galaxy, from 8$\deg$ to 45$\deg$ for an Sb galaxy and from
10$\deg$ to 60$\deg$ for an Sc galaxy.
                                  
In Figure \ref{periodic_Sa}, we see how for smaller pitch angles $(i
\lesssim 15\deg)$, the density response maxima coincides with the
imposed spiral arms potential. This means that the filled squares in the figure,
that represent the places in the arm where stars would crowd for long
times, settle down along the locus of the imposed spiral, making the
existence of stable long-lasting spiral arms more likely. On the
other hand, for spiral arms with pitch angles larger than $\sim 15
\deg$, the response maxima lag behind systematically the imposed spiral
arm potential, i. e., the pitch angles corresponding to the density response are smaller than the imposed ones.

Figure \ref{periodic_Sb} shows a similar behavior than the one for Sa
galaxies, but in this case, the pitch angle range is slightly wider.
For smaller pitch angles $(i \lesssim 18\deg)$ the density response supports
closely the imposed spiral arms. For $i > 18\deg$, the
density response produces spiral arms with smaller pitch angles than
the imposed locus, avoiding the support to the spiral structure.

Figure \ref{periodic_Sc} shows a similar behavior than Sa and Sb galaxies, but in this case, the pitch angle range is wider. For smaller pitch angles $(i \lesssim 20\deg)$ the density
response seems to support the imposed spiral arms. For $i >
20\deg$, the density response produces spiral arms with smaller pitch
angles than the imposed locus, avoiding to support the spiral
structure.

\begin{figure}
\includegraphics[width=1\textwidth]{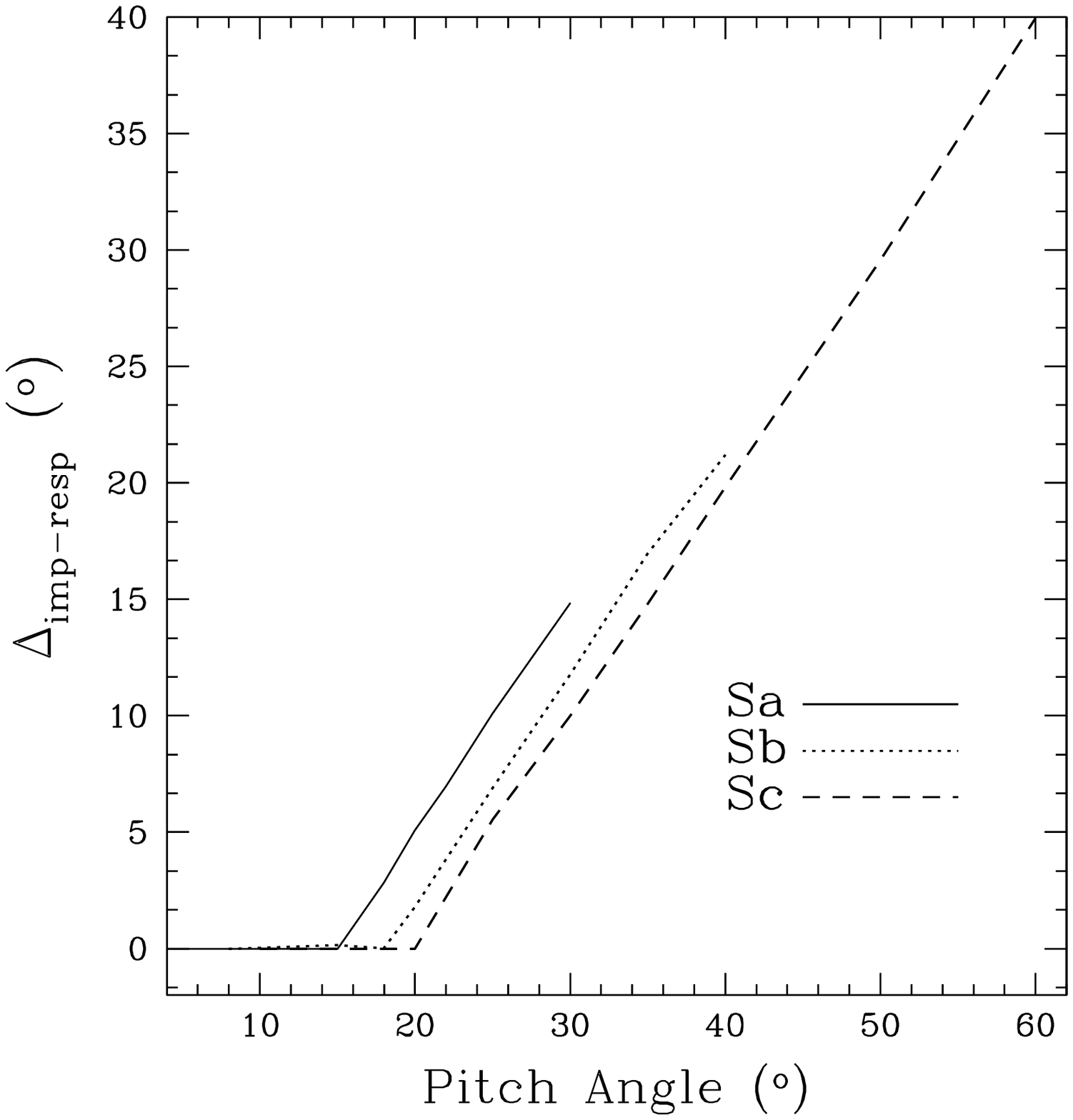}
\caption{	$\Delta_{imp-resp}$ is the phase angular difference between the density response and the imposed spiral potential. The solid line shows $\Delta_{imp-resp}$ for an Sa galaxy, the dotted line corresponds an Sb galaxy and the dashed line to an Sc galaxy. }
\label{diff}
\end{figure}

In the three types of galaxies, with pitch angles beyond 15$\deg$, 18$\deg$
and 20$\deg$ for Sa, Sb, and Sc galaxies, respectively, the density
response seems to ``avoid'' open long-lasting spiral arms. Long-lasting
spiral arms are not supported anymore after these limits; spiral arms
in these cases may be rather transient structures. In Figure \ref{diff} we show a plot of phase angular difference between the imposed spiral potential and the density response $vs.$ pitch angle of the imposed spiral arms.

\begin{figure}[H]
\includegraphics[width=1\textwidth]{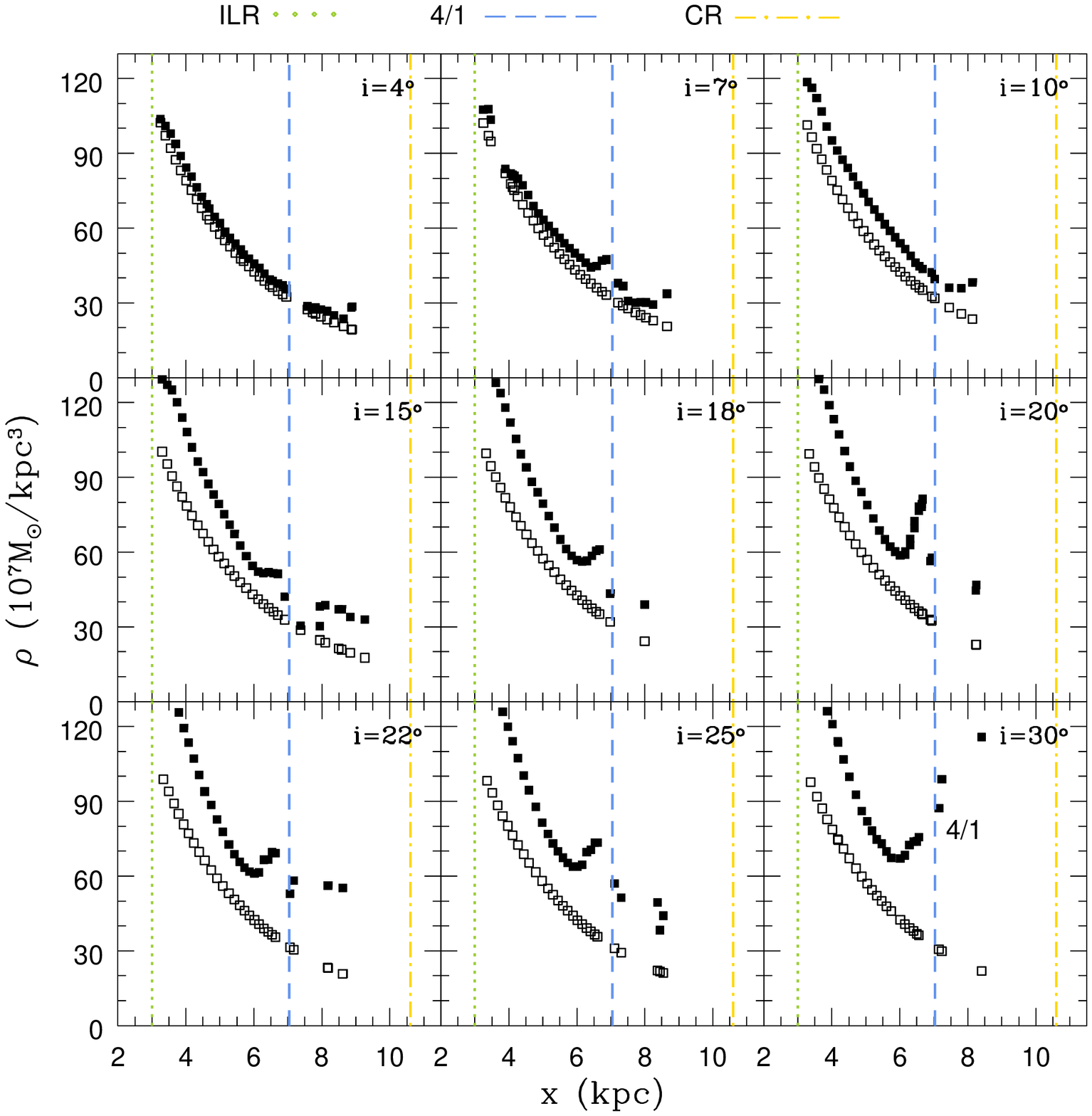}
\caption{In this figure, filled squares represent the density response
  of the spiral arms for an Sa galaxy, and open squares represent the
  imposed density, with pitch angles ranging from $4\deg$ to
  $30\deg$. The dotted, dashed and dotted-dashed lines show the inner Linblad resonance (ILR) position, 4/1 resonance position and corotation resonance (CR) position, respectively.}
\label{Response_Sa}
\end{figure}

\begin{figure}[H]
\includegraphics[width=1\textwidth]{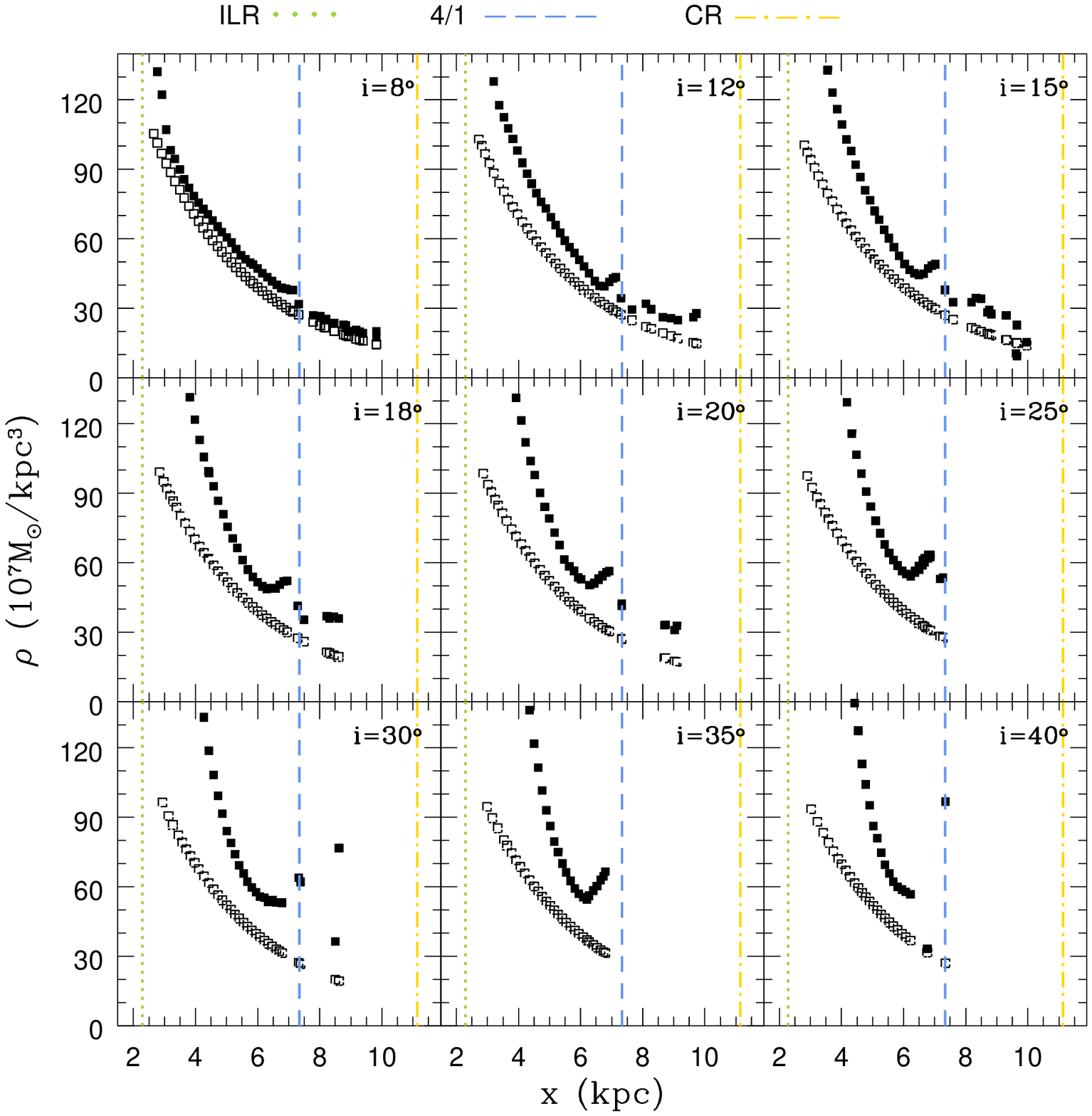}
\caption{In this figure, filled squares represent the density response
  of the spiral arms for an Sb galaxy, and open squares represent the
  imposed density, with pitch angles ranging from $8\deg$ to
  $40\deg$. The dotted, dashed and dotted-dashed lines show the inner Linblad resonance (ILR) position, 4/1 resonance position and corotation resonance (CR) position, respectively.}
\label{Response_Sb}
\end{figure}

\begin{figure}[H]
\includegraphics[width=1\textwidth]{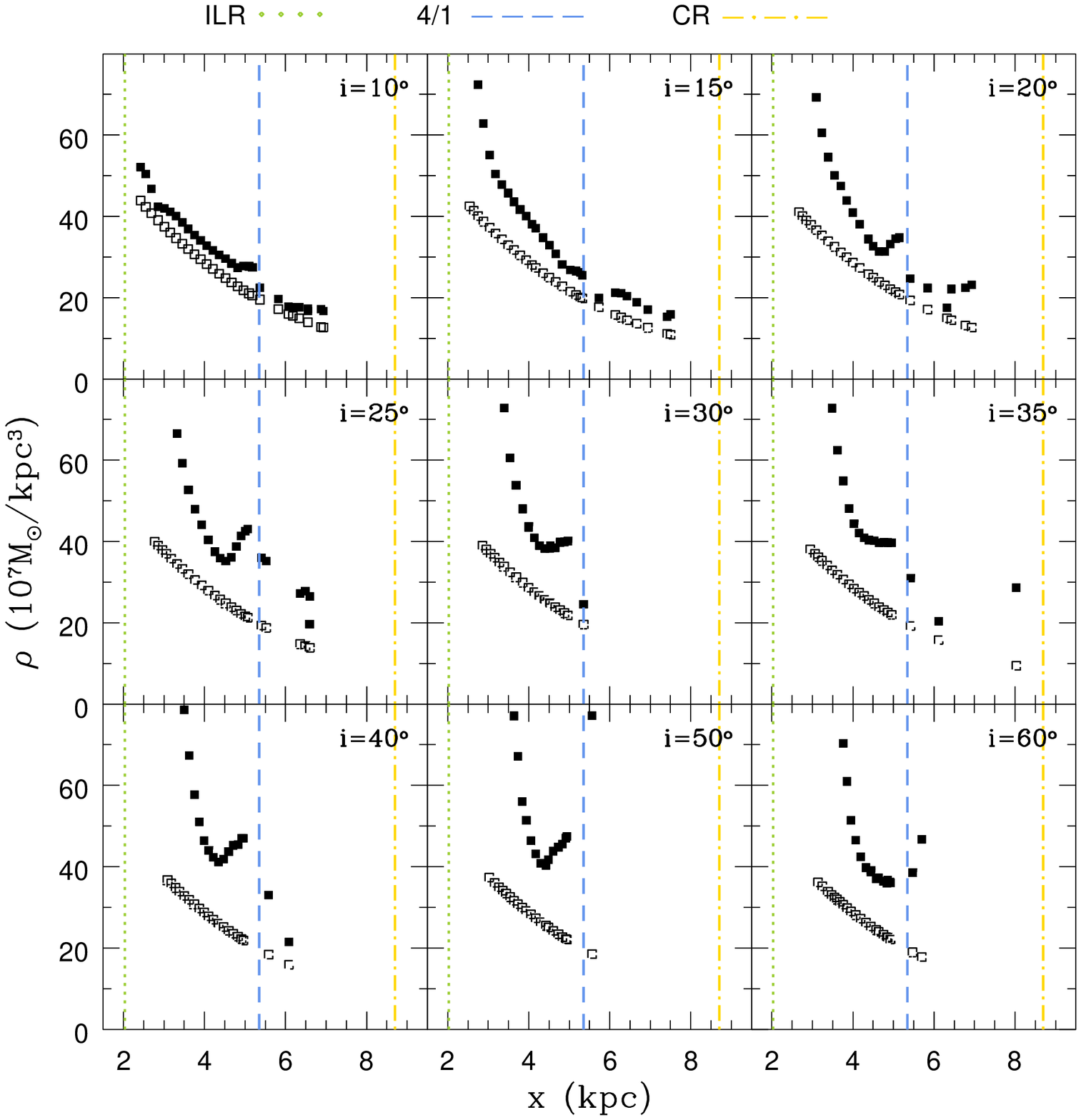}
\caption{In this figure, filled squares represent the density response
  of the spiral arms for an Sc galaxy, and open squares represent the
  imposed density, with pitch angles ranging from $10\deg$ to
  $60\deg$. The dotted, dashed and dotted-dashed lines show the inner Linblad resonance (ILR) position, 4/1 resonance position and corotation resonance (CR) position, respectively. }
\label{Response_Sc}
\end{figure}

An additional method to complement and to reinforce the results given
by periodic orbits, is the comparison of the spiral arms density
response (filled squares in Figures \ref{Response_Sa} --
\ref{Response_Sc}) with the imposed density (open squares in Figures
\ref{Response_Sa} -- \ref{Response_Sc}). In Figure \ref{Response_Sa},
we present the densities (the spiral arms density response, and the
spiral arms imposed density, i.e. PERLAS), for an Sa galaxy. As the density response maxima in the previous diagrams, this figure shows that for pitch angles up to $\sim 15\deg$, the density
response fits well with the imposed density. In Figure
\ref{Response_Sb}, we present the densities for an Sb galaxy. In this
figure we see almost the same behavior than in Figure
\ref{Response_Sa}, but in this case, the pitch angle limit, where the
density response does not fit the imposed arm, is $\sim 18\deg$. Finally,
in Figure \ref{Response_Sc}, we present the same diagrams for an Sc
galaxy. Here, the behavior is very similar to the Sa and Sb
cases, but the limit for the pitch angle, where the density response
is fitted to the imposed density, is $\sim 20\deg$.
 
With periodic orbits, maxima density response, and comparing the
density response and imposed density, in order to obtain support to
long-lasting spiral arms, the pitch angle should be smaller than
$\sim$ 15$\deg$, 18$\deg$ and 20$\deg$, for Sa, Sb and Sc galaxies, respectively. However, spiral arms evidently exist with larger pitch angles in galaxies, we propose then that these are rather in a transient form.

\subsection{Pitch Angle Restriction Based on Chaotic Motion} \label{chaos}

Confined chaotic orbits are able to support large-scale structures, such as spiral arms (Patsis \& Kalapotharakos 2011; Kaufmann \& Contopoulos 1996; Contopoulos \& Grosb\o l
1986). However, grand design structures are not expected to arise from systems where chaos fully dominates (Voglis \et 2006). In this study we found a restriction
based on chaotic behavior. To do this, we have produced an extensive
study of the Jacobi energy families in the phase space, as a function of the pitch angle in normal spiral galaxies. In this section we show that there is a limit to the pitch angle, for which chaos
becomes pervasive destroying all periodic orbits and the ordered orbits surrounding them, in the relevant spiral arm region.

\begin{figure}[H]
\includegraphics[width=1\textwidth]{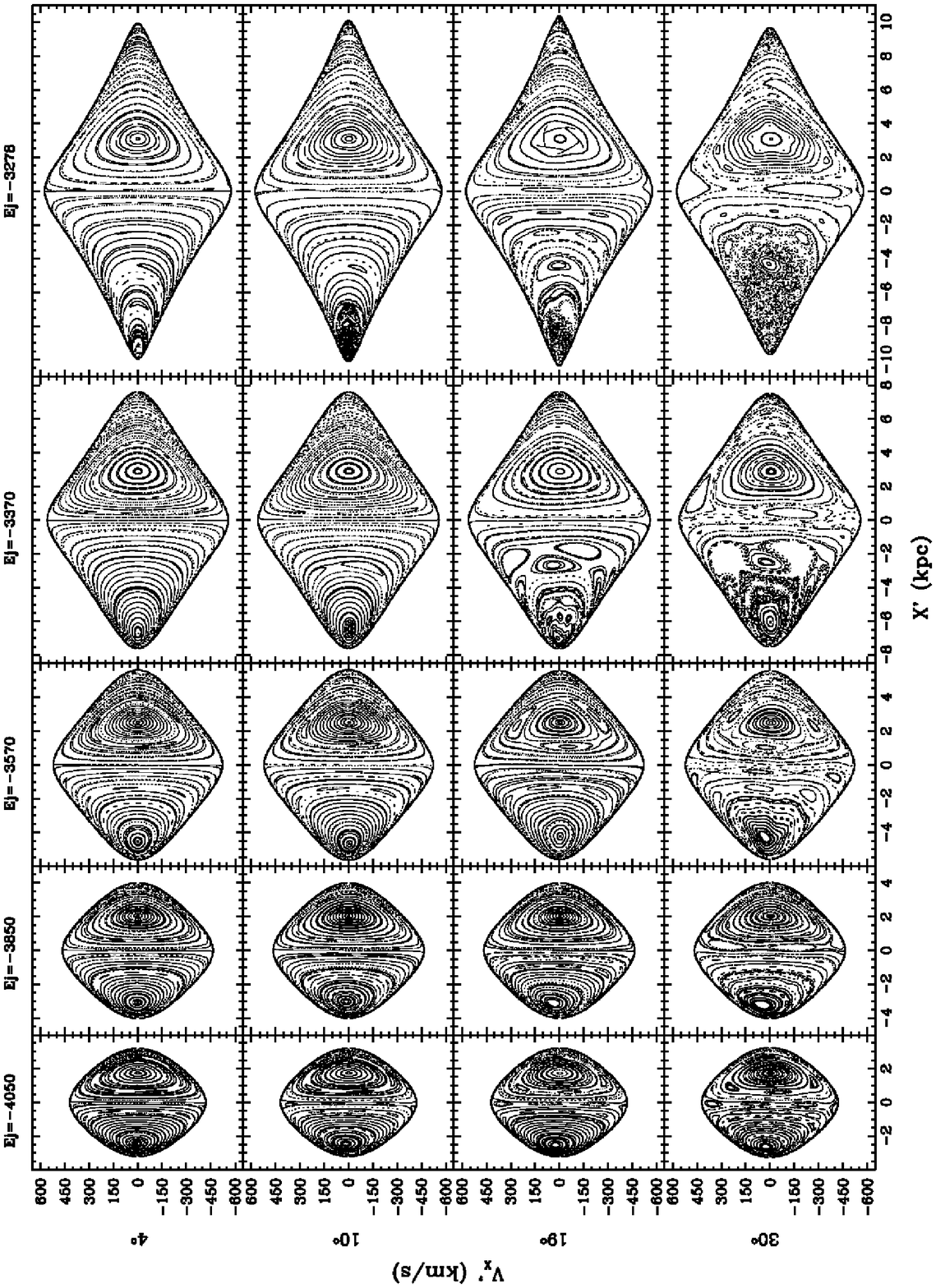}
\caption{Phase space diagrams for an Sa galaxy, with $E_J=[-4050,
    -3278]$, in units of $10^2 \, \rm {km}^2 \, \rm {s}^{-2}$. From
  top to bottom panels of the diagram, pitch angles go from 4$\deg$ to
  30$\deg$.}
\label{PS_Sa}
\end{figure}

\begin{figure}[H]
\includegraphics[width=1\textwidth]{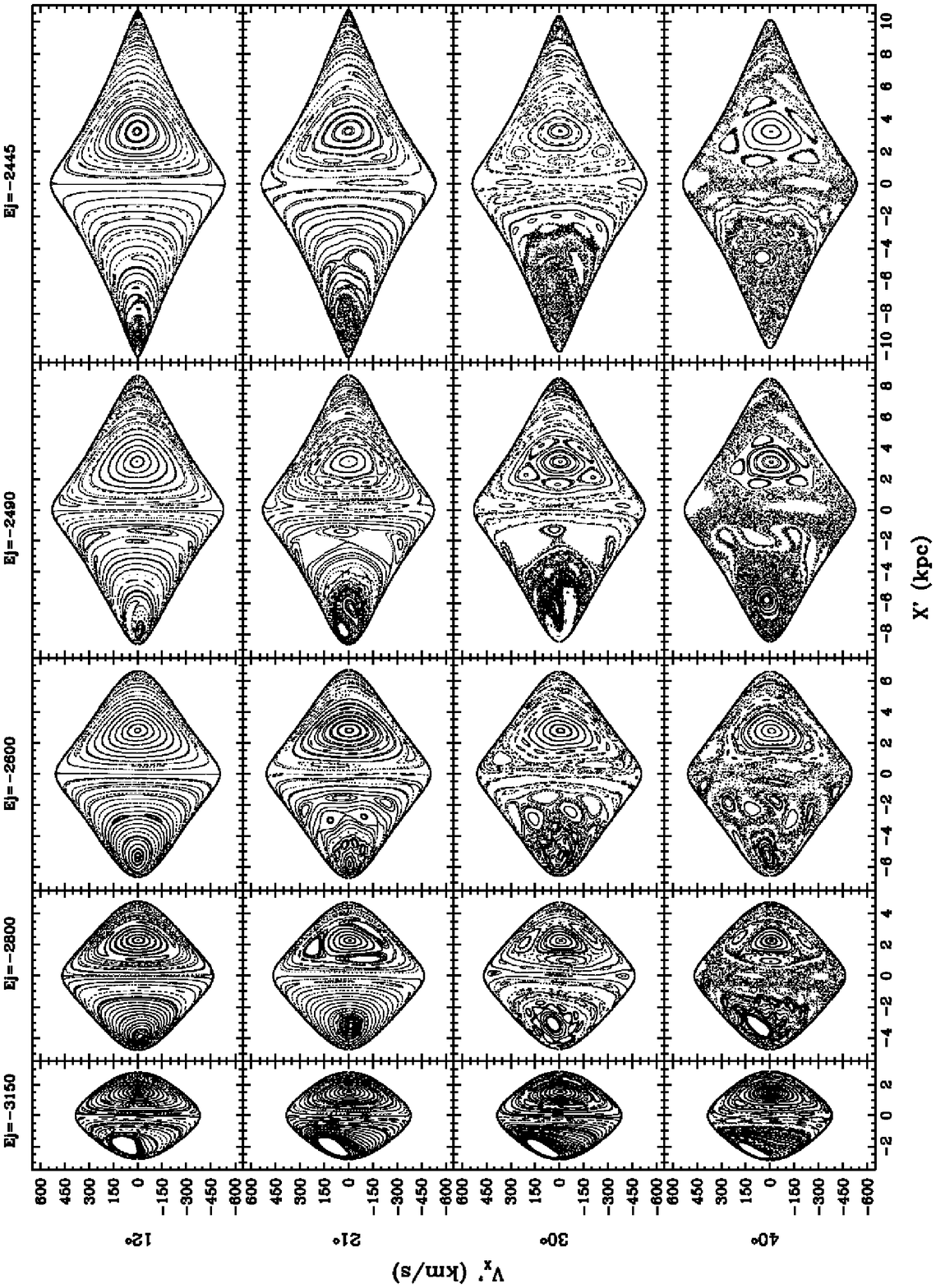}
\caption{Phase space diagrams for an Sb galaxy, with $E_J=[-3150,
    -2445]$, in units of $10^2 \, \rm {km}^2 \, \rm {s}^{-2}$. From
  top to bottom panels of the diagram, pitch angles go from 12$\deg$
  to 40$\deg$.}
\label{PS_Sb}
\end{figure}

\begin{figure}[H]
\includegraphics[width=1\textwidth]{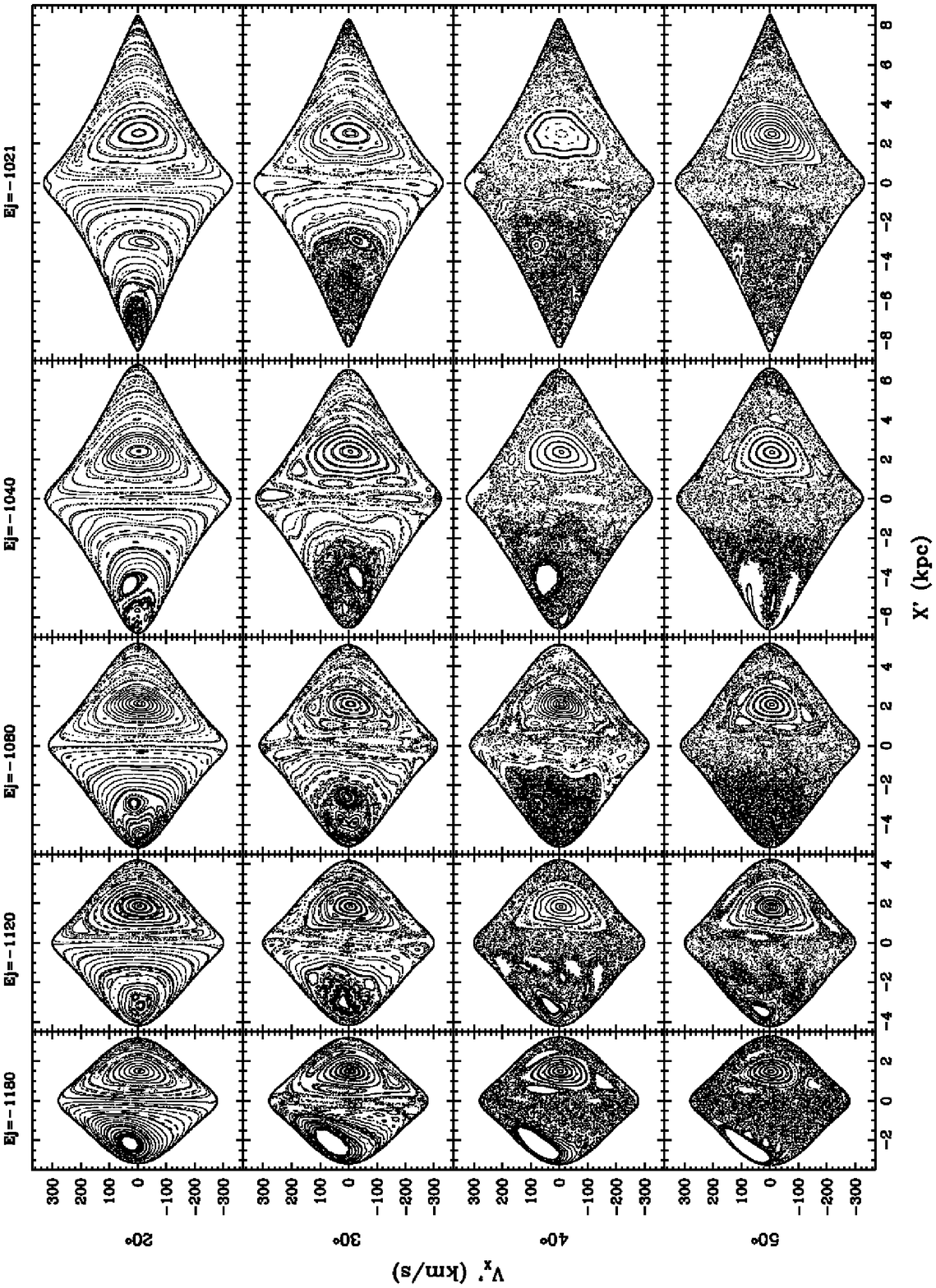}
\caption{Phase space diagrams for an Sc galaxy, with $E_J=[-1180,
    -1021]$, in units of $10^2 \, \rm {km}^2 \, \rm {s}^{-2}$. From
  top to bottom panels, pitch angles go from 20$\deg$ to 50$\deg$.}
\label{PS_Sc}
\end{figure}

We present here a set of phase space diagrams for each morphological
type (Figures \ref{PS_Sa} -- \ref{PS_Sc}). As in the study of periodic
orbits (Section \ref{order}), we employed an axisymmetric background
potential for an Sa, Sb and Sc galaxy, based on the parameters given
in Table \ref{tab:parameters}. In these experiments we have only
varied the pitch angles. Each mosaic has 20 panels, that show
phase space diagrams with different Jacobi constant families ranging from
$E_J=-4050$ to $-3278$ $\times 10^2 \, \rm {km}^2 \, \rm {s}^{-2}$ for
an Sa galaxy (where $E_J=-4100$, $-3410$, $-3275$ $\times 10^2 \, \rm {km}^2
\, \rm {s}^{-2}$ correspond approximately to the positions of ILR, 4/1 and CR resonances, respectively); from $E_J=-3150$ to $-2445$ $\times 10^2 \, \rm {km}^2
\, \rm {s}^{-2}$ for an Sb galaxy (where $E_J=-3290$, $-2550$, $-2442$ $\times 10^2 \, \rm {km}^2
\, \rm {s}^{-2}$ correspond approximately to the positions of ILR, 4/1 and CR resonances, respectively), and from $E_J=-1080$ to $-1021$
$\times 10^2 \, \rm {km}^2 \, \rm {s}^{-2}$ for an Sc galaxy (where $E_J=-1280$, $-1075$, $-1022$ $\times 10^2 \, \rm {km}^2
\, \rm {s}^{-2}$ correspond approximately to the positions of  ILR, 4/1 and CR resonances, respectively), covering
the total extension of spiral arms. 

Figure \ref{PS_Sa} shows Poincar\'e diagrams for an Sa galaxy, going
from 4$\deg$ (top line of diagrams) to 30$\deg$ (bottom line of
diagrams). For pitch angles between 4$\deg$ and 10$\deg$ (first two
lines of diagrams), the ordered orbits are dominating and simple,
periodic orbits support spiral arms up to corotation,
approximately. The onset of chaos is clear at about 10$\deg$. As a function of the pitch angle, for 19$\deg$ (third line of diagrams), the
orbital behavior is more complex, and it presents resonant
islands. For 30$\deg$ (bottom line of diagrams), the chaotic region
covers most of the regular prograde orbits. For pitch angles
beyond $\sim 30\deg$, chaos destroys periodic orbits.

Figure \ref{PS_Sb} shows Poincar\'e diagrams for an Sb galaxy, going
from 12$\deg$ (top line of diagrams) to 40$\deg$ (bottom line of
diagrams). For pitch angles of 10$\deg$ or less, the ordered orbits
dominate. As we increase the pitch angles, at 12$\deg$ (first line of
diagrams), the orbital behavior is mainly ordered, but close to
corotation there is already a small chaotic region. For 21$\deg$
(second line of diagrams), the onset of chaos is clear, and there is a
variety of complicate orbits. For 30$\deg$ (third line of diagrams),
the chaotic region increases, and the orbital structure becomes much
more complex. For 40$\deg$ (bottom line of diagrams), the chaotic
region covers almost all regular prograde orbits. For pitch angles
beyond $\sim 40\deg$ chaos destroys periodic orbits.

In Figure \ref{PS_Sc}, we present Poincar\'e diagrams for an Sc
galaxy, going from 20$\deg$ (top line of diagrams) to 50$\deg$ (bottom
line of diagrams). For a pitch angle of 20$\deg$ (first line of
diagrams) or smaller, orbits are always ordered and quite simple, the
spiral arms are supported up to corotation. For a pitch
angle of 30$\deg$, the orbital behavior is
slightly more complex, it shows some resonant islands and the chaotic region becomes
already important, but this is still contained by stable periodic
orbits. For 40$\deg$, chaos dominates the region around the stable periodic orbits. For 50$\deg$ (bottom line of diagrams), chaos covers practically all the regular prograde region, very close to the main periodic orbits. For pitch angles larger than
$\sim$50$\deg$  periodic orbits are destroyed by fully developed chaos. It might be natural wondering why for bars it is ``permitted'' to have a 90$\deg$ pitch angle, although we are stating here it is not possible for spiral arms. However, comparing directly the bar and the spiral arms regions is not straightforward, since the density structure and dynamics of the regions where they belong are quite different. 

The onset of chaos begins in the prograde region of Poincar\'e diagrams. The main
cause for chaotic motion has been attributed to resonance interactions (Contopoulos 1967; Martinet 1974; Athanassoula \et 1983; Pichardo \et 2003). In the case of
retrograde orbital regions, resonances are more widely separated than in the progade resonances case, which may explain why the onset and well developed chaos takes place in the prograde region first (and in general, only). However, it is worth noticing that in our experiments,  we have mantained the same
pattern angular speeds, that places resonances, at the same radii, increasing only pitch
angles,  and we have noticed that chaos (not originated only by the resonances position), increases dramatically. This means that at a given position of the resonances (posed by the angular speeds), increasing pitch angles, will widen the chaotic regions.


\section{Discussion and Conclusions}  \label{conclusions}

We have produced a set of models for observationally motivated
potentials to simulate typical Sa, Sb and Sc spiral galaxies with
three dimensional bisymmetric spiral arms. Observed galaxies,
classified as Sa, Sb or Sc, present a wide scatter in pitch angles
going from $\sim$4$\deg$ to 25$\deg$ for Sa galaxies, from
$\sim$8$\deg$ to 35$\deg$ for Sb galaxies and from $\sim$10$\deg$ to
50$\deg$ for Sc galaxies. With our models, we studied extensively the stellar
dynamical effects of the spiral arms pitch angle on the plane of the disk, ranging from 4$\deg$ to 40$\deg$, for an Sa galaxy, from
8$\deg$ to 45$\deg$, for an Sb galaxy and from 10$\deg$ to 60$\deg$,
for an Sc galaxy.

We found two important restrictions to the pitch angle. The first
restriction is based on the orbital ordered behavior. With the study
of periodic orbits and density response, we found that there is an
abrupt limit for the density response at approximately 15$\deg$ for Sa
galaxies, 18$\deg$ for Sb galaxies and 20$\deg$ for Sc galaxies. This limit denotes the end of orbital support of the density response to the imposed
spiral arm potential. In cases, where the spiral arms
potential is followed by their density response produced by
periodic orbital crowding, the spiral arms are more stable and could
be explained better as a long-lasting feature. Beyond these limits, the density response
produces systematically smaller pitch angles than the imposed spiral
arms. Galaxies with spiral arms beyond these limits would rather be
explained as transient features. 

We found a second restriction, this time based on chaotic orbital
behavior. In this case, we find a limit for the very same existence of spiral arms for each morphological type: for an Sa, the limit is $\sim 30\deg$, for an Sb, the limit is
$\sim40\deg$, and for an Sc the limit is $\sim50\deg$, for which spiral arms are so open that chaos dominates large regions of phase space, but there exist periodic
orbits supporting spiral arms. Beyond these limits for
each type of galaxy, chaos becomes pervasive destroying all orbital
support. 

With this orbital study, we are able to pose both a limit for steady long-lasting spiral 
arms, beyond which spiral arms are better explained as transient features and a limit for
maximum pitch angles (no matter their nature) in normal spiral galaxies before the system
becomes completely chaotic.

Several structural and dynamical parameters may play an important role in the stellar dynamics behavior, such as the pattern speed, the strength of the spiral arms (spiral arms mass), the density decay along the spiral arms, the axisymmetric components mass ratios, etc. In this paper we have isolated the effect of the pitch angle on stellar dynamics to understand at what extent we could impose restrictions to its properties in models for spiral galaxies based on order and chaos. We employ for this purpose, realistic values for the rest of the parameters that identify approximately typical Sa, Sb and Sc galaxies. A comprehensive study for other parameters is being prepared in an forthcoming paper. We provide here preliminary results on those studies for three of the relevant parameters within the observational or theoretical values for spiral galaxies: pattern speed, spiral arms mass and density decay of the spiral arms. For the pattern speed, although the construction and radial extension of periodic orbits (length of the spiral arms) depends sensibly on the pattern speed, the existence of the periodic orbits supporting the spiral structure inside the corotation resonance (or 4/1 resonance) seems not too sensitive, within the observed (20 to 30 km s$^{-1}$ kpc$^{-1}$) limits for real spiral galaxies. This means that, within the observed limits for spiral angular speeds in galaxies, the effect on the restriction of the angular speed on whether spiral arms are transient or long lived structures seems to be much less important. The spiral to disk mass ratio, as well as the angular speed, is a very restricted quantity by observations and theory. Translating the amplitudes or relative forcing, to spiral arms mass, the masses to obtain self-consistent models go upto 5 or 6\% of the mass of the disk. As long as we change within these limits the spiral arms mass, the restriction to the pitch angle for the spiral arms to be supported by periodic orbits is only slightly sensitive. Regarding the density fall of the spiral arms we studied several types of both lineal decay and exponential decay. This specific parameter results of little consequence as long as the linear decay slope goes approximately similar to the average slope of the exponential fall in the first half of the spiral arms extent. In a future work, we will present a detailed stellar dynamical study with these and other parameters of spiral galaxies.

\acknowledgments

We acknowledge the anonymous referee for enlightening comments that improved this work. We thank DGAPA-PAPIIT through grants IN110711 and IN112911.


\begin{thebibliography}   

\bibitem[Allen \& Santill\'an (1991)]{AS91} 
Allen, C. \& Santill\'an, A. 1991, \rmxaa, 22, 256

\bibitem[Antoja et al. (2009)]{A09}
Antoja, T., Valenzuela, O., Pichardo, B., Moreno, E.,
Figueras, F. \& Fern\'andez D. 2009, \apj, 700, L78 

\bibitem[Antoja et al.(2011)]{2011MNRAS.418.1423A} 
Antoja, T., Figueras, F., Romero-G{\'o}mez, M., et al.\ 2011, \mnras, 418, 1423

 \bibitem[Athanassoula et al., 1983] {Ath83}
 Athanassoula, E., Bienaym\'e, O., Martinet, L., \& Pfenniger, D. 1983, A\&A, 127, 349 

\bibitem[Block \et (2002)]{B02} 
Block, D. L., Bournaud, F., Combes, F., Puerari, I. \& Buta, R. 2002,\aap, 394, L35

\bibitem[Block et al.(2004)]{2004AJ....128..183B} 
Block, D.~L., Buta, R., Knapen, J.~H., et al.\ 2004, \aj, 128, 183

\bibitem[Binney \& Tremaine(1994)]{BT94}
Binney, J. \& Tremaine, S. 1994, {\it Galactic Dynamics}(Pinceton University Press)

\bibitem[Broshe (1971)]{B71}
Brosche, P. 1971, \aap, 13, 293

\bibitem[Buta \& Block (2001)]{BB01}
Buta, R. \& Block, D. L. 2001, \apj, 550, 243 

\bibitem[Buta et al.(2004)]{2004AJ....127..279B} 
Buta, R., Laurikainen, E., \& Salo, H.\ 2004, \aj, 127, 279

\bibitem[Buta \et  (2005)]{B05}
Buta, R., Vasylev, S., Salo, H. \& Laurikainen, E. 2005, AJ, 130, 523

\bibitem[Combes \& Sanders (1981)]{CS81}
Combes, F. \& Sanders, R. H. 1981, A\&A, 96, 164

\bibitem[Contopoulos  (1967)]{C67}
Contopoulos, G. 1967, Bull. Astron. (Ser. 3), 2, 223

\bibitem[Contopoulos  (1983)]{C83}
Contopoulos, G. 1983, \aap, 117, 89

\bibitem[Contopoulos  (1995)]{C95}
Contopoulos, G. 1995, NYASA, 751, 112

\bibitem[Contopoulos(2002)]{C02}
Contopoulos, G. 2002, {\it Order and Chaos in Dynamical Astronomy}, Springer, NY
 
\bibitem[Contopoulos \& Grosb\o l(1986)]{CG86}
Contopoulos, G. \& Grosb\o l, P. 1986, A\&A, 155, 11

\bibitem[Contopoulos \& Grosbol(1988)]{1988A&A...197...83C} 
Contopoulos, G., \& Grosbol, P.\ 1988, \aap, 197, 83 

\bibitem[Contopoulos \& Patsis (2008)]{CP08}
Contopoulos, G. \& Patsis, P. A., eds., 2008, Chaos in Astronomy,
Springer, Berlin

\bibitem[Contopoulos, Varvoglis \& Barbanis (1987)]{CVB87}
Contopoulos, G., Varvoglis, H. \& Barbanis, B. 1987,
\aap, 172, 55

\bibitem[Danver (1942)]{D42}
Danver, C. G. 1942, Ann. Lund. Obs., 10, 162

\bibitem[Davis et al.(2012)]{2012ApJS..199...33D} Davis, B.~L., Berrier, 
J.~C., Shields, D.~W., et al.\ 2012, \apjs, 199, 33 

\bibitem[de Vaucouleurs (1959)]{dV59}
de Vaucouleurs G. 1959, Hanbuch der Physik, 53, 275

\bibitem[Drimmel  (2000)]{D00}
Drimmel, R. 2000, \aap, 358, L13

\bibitem[Egusa et al.(2009)]{2009ApJ...697.1870E} 
Egusa, F., Kohno, K., Sofue, Y., Nakanishi, H., \& Komugi, S.\ 2009, \apj, 697, 1870 

\bibitem[Fathi et al. (2009)]{F09}
Fathi, K., Beckman, J. E., Pi\~nol-Ferrer, N., Hern\'andez, O.,
Mart\'inez-Valpuesta, I. \& Carignan, C. 2009, \apj, 704, 1657 

\bibitem[Foyle et al.(2011)]{2011ApJ...735..101F} Foyle, K., Rix, H.-W., 
Dobbs, C.~L., Leroy, A.~K., \& Walter, F.\ 2011, \apj, 735, 101 

\bibitem[Fujii \& Baba(2012)]{2012MNRAS.427L..16F} Fujii, M.~S., \& Baba, J.\ 2012, \mnras, 427, L16 

\bibitem[Goldreich \& Lynden-Bell(1965)]{1965MNRAS.130..125G} Goldreich, P., \& Lynden-Bell, D.\ 1965, \mnras, 130, 125 

\bibitem[Groot (1925]{G25}
Groot, H. 1925, MNRAS, 85, 535

\bibitem[Grosb\o l (2003)]{G03} 
Grosb\o l, P. 2003, LNP, 626, 201

\bibitem[Grosb{\o}l \& Dottori(2009)]{2009A&A...499L..21G}
Grosb{\o}l, P., \& Dottori, H.\ 2009, \aap, 499, L21 

\bibitem[Grosb\o l, \& Patsis (1998)]{GP98} 
Grosbol, P. J. \& Patsis, P. A. 1998, \aap, 336, 840

\bibitem[Grosb{\o}l et al.(2002)]{2002ASPC..275..305G} 
Grosb{\o}l, P., Pompei, E., \& Patsis, P.~A.\ 2002, Disks of Galaxies: Kinematics,
  Dynamics and Peturbations, 275, 305

\bibitem[Holmberg (1958)]{H58}
Holmberg, E. 1958, Medd. Lund. Ast. Obs. Ser. II, No. 136

\bibitem[Hubble (1926)]{H26} 
Hubble, E. P. 1926, ApJ, 64,321

\bibitem[Hubble (1936)]{H36} 
Hubble, E. P. 1936, {\it The Realm of the Nebulae.} New Haven: Yale Univ. Press

\bibitem[Julian \& Toomre(1966)]{1966ApJ...146..810J} Julian, W.~H., \& Toomre, A.\ 1966, \apj, 146, 810 

\bibitem[Kalapotharakos et al. (2010)]{K10}
Kalapotharakos, C., Patsis, P. A. \& Grosb\o l, P. 2010, \mnras, 403, 83

\bibitem[Kaufmann \& Contopoulos(1996)]{1996A&A...309..381K} 
Kaufmann, D.~E., \& Contopoulos, G.\ 1996, \aap, 309, 381 


\bibitem[Kawata et al.(2012)]{2012EPJWC..1907006K} Kawata, D., Grand, R.~J.~J., \& Cropper, M.\ 2012, European Physical Journal Web of Conferences, 19, 7006 

\bibitem[Kennicutt (1981)]{K81} 
Kennicutt, R. 1981, AJ, 86, 1847

\bibitem[Laurikainen et al.(2004)]{2004MNRAS.355.1251L} 
Laurikainen, E., Salo, H., Buta, R., \& Vasylyev, S.\ 2004, \mnras, 355, 1251

\bibitem[Laurikainen \& Salo (2002)]{LS02}
Laurikainen, Eija \& Salo, Heikki 2002, \mnras, 337, 1118

\bibitem[Lin \& Shu (1964)]{LS64}
Lin, C. C. \& Shu, F. H. 1964, Apj, 140, 646

\bibitem[Ma et al. (2000) ]{MA00}
Ma, Jhun, Zhao, Jun-liang, Zhang, Fei-peng, \& Peng, Qiu-he
2000, Chinese Astronomy and Astrophysics, 24, 435

\bibitem[Ma (2002)]{2002A&A...388..389M} 
Ma, J.\ 2002, \aap, 388, 389 

\bibitem[Martinet (1974)]{Mart74}
Martinet, L. 1974, A\&A, 32, 329

\bibitem[Martos et al. (2004) ]{MH04}
Martos, M., Hernandez, X., Y\'a\~nez, M., Moreno, E. \& Pichardo, B.
2004, MNRAS, 350, 47

\bibitem[Maupome, Pi\c smi\c s \& Aguilar (1981)]{MPA81} Maupom\'e,
  L., Pi\c smi\c s, P. \& Aguilar, L. 1981, Rev. Mexicana
  Astron. Astrof., 6, 45

\bibitem[Miyamoto \& Nagai (1975)]{MN75}
Miyamoyo, M., \& Nagai, R. 1975, Pub. Astr. Soc. Japan, 27, 533

\bibitem[Patsis (2008)]{P08}
Patsis, P. A. 2008, AN, 329, 930

\bibitem[Patsis, Contopoulos \& Grosb\o l (1991)]{PCG91}
Patsis, P. A., Contopoulos, G. \& Grosb\o l P. 1991, A\&A, 243, 373

\bibitem[Patsis \& Kalapotharakos(2011)]{2011MSAIS..18...83P} Patsis,
  P.~A., \& Kalapotharakos, C.\ 2011, Memorie della Societa
  Astronomica Italiana Supplementi, 18, 83

\bibitem[Patsis, Kaufmann et al. (2009)]{PKGB09} 
Patsis, P. A., Kaufmann, D. E., Gottesman, S. T. \& 
Boonyasait, V. 2009, \mnras, 394, 142
 
 \bibitem[P\'erez-Villegas et al. 2012] {PV12}
 P\'erez-Villegas, A., Pichardo, B., Moreno, E., Peimbert, A., \& Vel\'azquez, H. M. 
 2012, ApJL, 745, L14
 
\bibitem[Pichardo, Martos, Moreno \&  Espresate (2003)]{PMME03}
Pichardo, B., Martos, M., Moreno, E. \& Espresate, J. 2003, ApJ, 582, 230

\bibitem[Pi\c smi\c s \& Maupom\'e (1975)]{PM75}
Pi\c smi\c s, P. \& Maupom\'e, L. 1978, Rev. Mexicana Astron. Astrof., 2, 319

\bibitem[Pizagno et al. (2005)]{Pizagno05}
Pizagno, J., Prada, F., Weinberg, D. H., Rix, H. W., Harbeck, D.,
Grebel, E. K., Bell, E. F., Brinkmann, J., Holtzman, J. \& West, A. 
2005, \apj, 633, 844

\bibitem[Roberts \& Haynes (1994)]{RH94}
Roberts, M. S. \& Haynes, M. P. 1994, ARA\&A, 32, 115

\bibitem[Roberts, Huntley \& van Albada (1979)]{RHV79}
Roberts, W.W., Jr., Huntley, J. M., \& van Albada, G. D. 1979, \apj, 233, 67

\bibitem[Rubin et al.(1985)]{1985ApJ...289...81R} Rubin, V.~C., Burstein, D., Ford, W.~K., Jr., \& Thonnard, N.\ 1985, \apj, 289, 81

\bibitem[Sage (1993)]{S93}
Sage, L. J. 1993, A\&A, 272, 123

\bibitem[Sandage (1961)]{S61}
Sandage A. 1961, The Hubble Atlas of Galaxies. Carnegie Institution of Washington, Washington D.C.

\bibitem[Sandage (1975)]{S75}
Sandage A. 1975, Galaxies and Universe-Stars and Stellar Systems Chicago: University of Chicago Press

\bibitem[Sandage(2000)]{2000PASP..112..504S} Sandage, A.\ 2000, \pasp, 112, 504 

\bibitem[Savchenko \& Reshetnikov(2011)]{2011AstL...37..817S} Savchenko, S.~S., \& Reshetnikov, V.~P.\ 2011, Astronomy Letters, 37, 817 

\bibitem[Seigar et al.(2008)]{2008ApJ...678L..93S} Seigar, M.~S., Kennefick, D., Kennefick, J., \& Lacy, C.~H.~S.\ 2008, \apjl, 678, L93 

\bibitem[Sellwood (2011)] {S11}
Sellwood, J. A. 2011, MNRAS, 410, 1637

\bibitem[Sellwood \& Carlberg(1984)]{1984ApJ...282...61S} Sellwood, J.~A., \& Carlberg, R.~G.\ 1984, \apj, 282, 61 

\bibitem[Sersic (1987)]{S87}
Sersic, J. L. 1987, Extragalactic Astronomy, trans. Li Zong-yun, rev. Huang Ke-liang. Beijing: Science Press

\bibitem[Shields et al.(2010)]{2010AIPC.1294..283S} Shields, D.~W., Hughes, J.~A., Barrows, S.~R., et al.\ 2010, American Institute of Physics Conference Series, 1294, 283

\bibitem[Sofue \& Rubin (2001)]{SR01}
Sofue, Y. \& Rubin, V. 2001, \araa, 39, 137 

\bibitem[Voglis et al. (2006)]{VSK06}
Voglis, N., Stavropoulos, I. \& Kalapotharakos, C. 2006, \mnras,
372, 901

\bibitem[Von der Pahlen (1911)]{VP11}
Von der Pahlen 1911, AN, 188, 249

\bibitem[Vorobyov (2006)]{V06}
Vorobyov, E. I. 2006, MNRAS, 370, 1046

\bibitem[Weinzirl et al. (2009)]{Weinzirl09}
Weinzirl, T., Jogee, S., Khochfar, S., Burket, A. \& Kormendy, J. 2009,
\apj, 696, 411

\bibitem[]{} 

\end{thebibliography}
\end{document}